\documentclass[aps,twocolumn,showpacs,amsmath,superscriptaddress,amssymb,prc]{revtex4-2}
\usepackage{graphicx,here}
\usepackage{multirow}
\usepackage{tikz}
\usepackage{epstopdf}
\usepackage[percent]{overpic}
\usepackage{scrextend}
\usepackage{xcolor,xspace}
\usepackage[ulem=normalem]{changes}
\usepackage{hyperref}
\hypersetup{colorlinks=True,urlcolor=blue,linkcolor=blue,citecolor=blue,filecolor=black}

\colorlet{Changes@Color}{red}
\usepackage{dcolumn,color,footnote,bm,braket}
\usepackage{url,longtable,tabularx}
\usepackage{threeparttable}

\usepackage{fancybox}
\usetikzlibrary{matrix}

\newcommand\+{\dagger}
\newcommand{\be}[4]{B(E2; #1^+_{#2} \!\to\! #3^+_{#4})}
\newcommand{\bej}[4]{#1^+_{#2} \!\to\! #3^+_{#4}}
\newcommand{\rhoe}[4]{\rho^2(E0; #1^+_{#2} \!\to\! #3^+_{#4})}

\begin{document}

\title{Effects of
shape coexistence and configuration mixing on
low-lying states in tellurium isotopes}

\author{Kosuke Nomura}
\email{nomura@sci.hokudai.ac.jp}
\affiliation{Department of Physics, 
Hokkaido University, Sapporo 060-0810, Japan}
\affiliation{Nuclear Reaction Data Center, 
Hokkaido University, Sapporo 060-0810, Japan}

\date{\today}

\begin{abstract}
Low-energy quadrupole collective states
in even-even tellurium (Te) isotopes are
studied using the interacting boson model
with configuration mixing.
The corresponding
Hamiltonian is determined by
means of the microscopic
nuclear structure calculations
within the self-consistent mean-field
method employing a given
energy density functional and
pairing interaction.
Calculated low-energy
levels for nonyrast states
show a parabolic behavior characteristic
of the shape-coexisting structure.
The intruder prolate-shape
configuration is shown to mix
strongly with the normal
oblate-shape configuration,
and play an important role
in determining the low-lying
structure in the Te isotopes
near the middle of
the neutron major shell closures.
\end{abstract}

\maketitle

\section{Introduction}

Low-energy structure of nuclei
near the proton major shell closure $Z=50$
has been considered a
textbook example exhibiting a vibrational
energy spectrum of multiphonon excitations.
However, in light of recent experimental studies
producing new spectroscopic data,
this interpretation does not
necessarily apply to many of those
nuclei in this mass region.
For cadmium (Cd) isotopes in the middle
of the neutron major shells $N=50$ and 82,
for instance,
some low-spin states were observed at
excitation energies close to those of
the two- or three-phonon multiplets;
these states cannot be explained
in terms of a pure vibrational picture.
These additional low-energy states in
the Cd isotopes have been attributed
to contributions of intruder
excitations across
the proton $Z=50$ major shell
\cite{federman77,heyde85,heyde1992,wood1992,heyde95,heyde2011}.
A similar mechanism is expected to
be at work in tin (Sn) isotopes.

Noticeable effects of the cross-shell
intruder excitations are the parabolic
behaviors of bands built on the
low-lying excited $0^+$ states.
These features are most
spectacularly seen in neutron-deficient
lead (Pb) and mercury (Hg) nuclei, which are
close to the proton $Z=82$ major shell.
The appearance of the low-energy
intruder $0^+$ band has been
interpreted as a signature
of coexistence of intrinsic shapes
\cite{heyde1992,wood1992,heyde2011,garrett2022,leoni2024}.
In $^{186}$Pb, for instance,
the ground-state band
has been empirically associated with
a spherical shape, and the
low-energy $0^+_2$ and $0^+_3$
excited bands
were attributed
to be a weakly oblate-deformed and
a strongly prolate-deformed shapes
\cite{andre00};
updated experimental data
for this nucleus
suggested that the
$0^+_2$ and $0^+_3$ bands
were associated
with the prolate and oblate
configurations, respectively
\cite{ojala2022-186Pb}.
In the mean-field approximation
\cite{bengtsson1987,bengtsson89,
wood1992,andre00,cwiok2005,heyde2011},
coexistence of multiple shapes
in the vicinity of the
ground state is inferred
from the presence of several
local minima
in the potential energy surface (PES)
in terms of relevant collective
coordinates.
The shape coexistence has been
observed in a wide range of the nuclear chart,
and is nowadays regarded
as a universal phenomenon in
nuclear many-body systems.

Here the low-energy structure
of even-even tellurium (Te)
isotopes is of special interest.
Since they are also close to
the proton $Z=50$ major shell,
similar cross-shell excitations
to those in the Cd and Sn isotopic chains
are expected to take place
and play relevant roles at low energy.
The effect of shape coexistence in
the mass region below the $Z=50$
major shell, including 
the midshell Cd and Sn isotopes
and those zirconium and molybdenum
nuclei near $N=60$,
has been extensively studied
experimentally, and by
related theoretical approaches
(see recent reviews
\cite{garrett2022,leoni2024}
and references therein).
Most recent studies
on nuclei near $Z=50$ include
ones that showed
appearance of multiple
shape coexistence in $^{110,112}$Cd
\cite{garrett2019,garrett2020}
from comparisons
to beyond-mean-field calculations,
and those for the neighboring
Sn isotopes in the neutron midshell
\cite{spieker2018-sn112114,
petrache2019-sn116,ortner2020-sn118},
which suggested possible
configuration mixing in these
isotopes on the basis of the
phenomenological interacting
boson model (IBM) calculations.
Concerning the Te isotopes,
earlier experimental study \cite{rikovska1989}
suggested intruder states in those
nuclei near the middle of the
neutron major shells.
Low-energy levels in
$^{116}$Te \cite{vonspee2024} and $^{118}$Te
\cite{mihai2011} were measured
in more recent studies,
in which the emergence of shape
coexistence was discussed.
Lifetime measurements were carried out
to extract the systematic of the
excitation energies
and $B(E2)$ values for low-lying
states
\cite{cederlof2023-118Te,cbli2024,pascu2025}. 
Related theoretical calculations
that were previously reported include
those within the nuclear shell model
\cite{qi2016,kaneko2021,sharma2022,cbli2024},
static \cite{sharma2019,bonatsos2022} 
and beyond \cite{prochniak1999,libert2007,delaroche2010,CEA,suzuki2024}
mean-field methods,
the IBM \cite{sambataro1982,
rikovska1989,Lehmann97,pascu2010,
mihai2011,gupta2023,vonspee2024},
the geometrical collective model
\cite{budaca2021}
and effective field theory
\cite{coelloparez2015}. 
The interpretation of the nuclear structure
in Te nuclei in terms of the shape coexistence
does not appear to have been
established as firmly as for
those nuclei below $Z=50$.
Further theoretical
investigations would help
identifying regions
of shape coexistence
in the neutron-deficient
$Z>50$ nuclei.

In the present work,
roles played by
the intruder configurations in
the low-lying structure of 
midneutron-shell Te isotopes
are investigated by using
the IBM in its appropriate version
that incorporates configuration
mixing of the normal and intruder states.
This version of the IBM
was previously applied
to identify possible intruder
states in Cd (e.g.,
\cite{heyde1982,heyde1992,
sambataro1982,deleze1993a,deleze1993b,
jolie1995,lehmann1995,heyde95,DeCoster96,
Lehmann97,garrett2007,garrett2008,nomura2018cd,
leviatan2018,gavrielov2023}),
Sn (e.g.,
\cite{spieker2018-sn112114,
petrache2019-sn116,ortner2020-sn118}),
and Te (e.g.,
\cite{sambataro1982,rikovska1989,heyde1992,
Lehmann97,pascu2010,mihai2011,vonspee2024})
nuclei.
In most of these studies,
the IBM calculations were performed
either with model parameters
being fitted to observed
low-energy spectroscopic properties
or assuming dynamical symmetries
for normal and intruder configurations.
In Ref.~\cite{deleze1993b},
a microscopic derivation of the
IBM parameters for the normal 
configuration of Cd nuclei
was also made
using the Otsuka-Arima-Iachello mapping
procedure \cite{OAI}.
The present study is based on the
microscopic framework of the
nuclear energy density functional (EDF):
the self-consistent mean-field
(SCMF) calculations employing
a given EDF provide microscopic inputs
to determine the parameters of
the IBM Hamiltonian that incorporates
the configuration mixing
\cite{nomura2012sc,nomura2016sc}.
This method was previously
employed for systematic studies
on the shape evolution
and coexistence, and related
spectroscopic properties
in different mass regions:
neutron-deficient Pb \cite{nomura2012sc}
and Hg \cite{nomura2013hg},
mid-neutron-shell Cd \cite{nomura2018cd},
and neutron-rich Sr, Zr, Mo, and Ru
nuclei near $N=60$ \cite{nomura2016zr},
neutron-deficient and neutron-rich
Ge, Se \cite{nomura2017ge},
and Kr \cite{nomura2017kr} isotopes.

For the microscopic part
of the theoretical analysis,
the SCMF calculations are performed
within the framework of the
relativistic Hartree-Bogoliubov (RHB)
model
\cite{vretenar2005,niksic2011,DIRHB,DIRHBspeedup}.
The RHB framework was actually
adopted in recent studies
\cite{sharma2019,suzuki2024}
in order to analyze
shape evolution and coexistence
in the Te isotopes.
The RHB self-consistent
calculations in these
studies produced the PESs,
which exhibit
a prolate-to-oblate shape transition
and coexistence in the midshell Te nuclei.
In Ref.~\cite{suzuki2024},
in particular,
spectroscopic properties of the
Te nuclei were computed in
terms of the quadrupole collective
Hamiltonian (QCH),
with the deformation-dependent
mass parameters and moments of
inertia determined by the
RHB-SCMF solutions.
The present work employs
the same microscopic inputs
from the RHB-SCMF calculations
as those for the QCH
considered in the
previous study \cite{suzuki2024},
and hence gives an alternative 
description of the
spectroscopic properties of Te isotopes
using the EDF-mapped IBM.

In the following,
the theoretical method is described
in Sec.~\ref{sec:model}.
The calculated results for the
PESs, low-energy spectra,
and electromagnetic transition
properties for the
even-even $^{108-126}$Te
are presented in Sec.~\ref{sec:results}.
Some dependence of the
calculated results on the
EDF is also discussed
in the same section.
Section~\ref{sec:summary} gives
a summary and conclusions.

\section{Theoretical framework\label{sec:model}}

The first step is a set of
constrained RHB-SCMF
calculations for each nucleus,
with the constraints being on the mass
quadrupole moments that are related
to the intrinsic axially symmetric
deformation $\beta$
and triaxiality $\gamma$ \cite{BM},
respectively.
The constrained calculations provide
PESs in terms of the $(\beta,\gamma)$
deformations.
The underlying interaction is
the density-dependent
point-coupling (DD-PC1) EDF
\cite{DDPC1} for the particle-hole part.
The DD-PC1 set is a representative
class of the relativistic EDF,
and has been successfully used
for numerous nuclear
structure studies.
For the particle-particle part,
a separable pairing
force of finite range of
Ref.~\cite{tian2009} is adopted,
with the standard value of the
pairing strength, 728 MeV fm$^3$,
throughout the present study.
Further details of the RHB calculations
are found in Ref.~\cite{suzuki2024}.

The second step is to construct
the boson Hamiltonian.
Here the neutron-proton version of the
IBM (referred to as IBM-2)
is employed, in which neutron
and proton degrees of freedom
are distinguished.
Accordingly,
the IBM-2 with configuration
mixing is hereafter denoted IBM2-CM.
The building blocks of the IBM-2
are neutron monopole $s_{\nu}$ and
quadrupole $d_{\nu}$ bosons, and
proton monopole $s_{\pi}$ and
quadrupole $d_{\pi}$ bosons.
The neutron (proton) $s_{\nu}$ and
$d_{\nu}$ ($s_{\pi}$ and $d_{\pi}$)
bosons reflect, from a microscopic point
of view \cite{OAIT,OAI},
collective monopole and quadrupole
pairs of valence neutrons (protons),
respectively.
The number
of neutron (proton) bosons,
denoted $N_{\nu}$ ($N_{\pi}$),
is equal to the number of neutron
(proton) pairs counted from the
nearest closed shell, and
is conserved for a given nucleus.
In the present case, $N_{\pi}=1$
and $N_{\nu}=3-8$
for $^{108-118}$Te, respectively,
and $N_{\nu}=7-4$ for $^{120-126}$Te,
respectively.

The intruder
configurations here represent
the $2n$-particle-$2n$-hole
($n=1,2,\ldots$) excitations
of protons across the $Z=50$ shell
and are described by the
unperturbed IBM-2 Hamiltonians
for $(N_{\nu},N_{\pi}+2n)$
bosons \cite{duval1981}.
As is usual
in previous IBM2-CM studies,
like-particle bosons
are not distinguished from
like-hole bosons.
In the present work,
two-boson configurations;
that is, normal (0p-0h)
and intruder (2p-2h), are considered.
This is based on the fact that,
as shown below,
the RHB-SCMF calculation
suggests basically two
(oblate global and prolate local)
minima in the PESs for
the midshell Te nuclei.
The IBM2-CM space is defined as a
direct sum of the two unperturbed IBM-2
spaces consisting of
$(N_{\nu},N_{\pi})$ and
$(N_{\nu},N_{\pi}+2)$ bosons,
and is expressed as
\begin{eqnarray}
 \label{eq:space}
[N_{\nu}\otimes N_{\pi}]
\oplus
[N_{\nu}\otimes (N_{\pi}+2)]
\; .
\end{eqnarray}
The corresponding IBM2-CM Hamiltonian
is written as
\begin{eqnarray}
 \label{eq:ham1}
\hat H
= 
\hat P_{1}
\hat H_{1}
\hat P_{1}
+
\hat P_{3}
(\hat H_{3} + \Delta)
\hat P_{3}
+
\hat V_{\rm mix} \; ,
\end{eqnarray}
where $\hat H_{1}$ and $\hat H_{3}$
denote the Hamiltonians for the
unperturbed 
spaces of $N_\pi=1$ and three bosons,
$\hat P_{1}$ and $\hat P_{3}$ are operators
projecting onto each unperturbed space,
$\Delta$ denotes the energy
required for the 2p-2h
excitation, and the last term stands
for the interaction that mixes
the two IBM-2 spaces.
For the unperturbed IBM-2 Hamiltonian
for the normal ($i=1$)
or intruder ($i=3$) configuration
the following simplified
form is adopted:
\begin{eqnarray}
\label{eq:ham2}
\hat H_{i}=
\epsilon_{d,i}
(\hat n_{d_{\nu}}
+\hat n_{d_{\pi}})
+
\kappa_{i}
\hat Q_{\nu,i}\cdot\hat Q_{\pi,i} \; ,
\end{eqnarray}
where in the first term
$\hat n_{d_{\rho}}=d^{\+}_{\rho}\cdot\tilde d_{\rho}$
represents the neutron ($\rho=\nu$)
or proton ($\rho=\pi$)
$d$-boson number operator
with the
energy $\epsilon_{d,i}$ with respect
to $s_{\rho}$ bosons, and the second term
represents the quadrupole-quadrupole
interaction between neutron and proton
bosons with strength $\kappa_{i}$.
The quadrupole operator
$\hat Q_{\rho,i}=
s_{\rho}^\+\tilde d_{\rho}
+d_{\rho}^\+\tilde s_{\rho}
+\chi_{\rho,i}(d_{\rho}^\+\times\tilde d_{\rho})^{(2)}$
with $\chi_{\rho,i}$ being
a dimensionless parameter.
The mixing term reads
\begin{eqnarray}
\label{eq:vmix}
 \hat V_{\rm mix}
=\omega(s_{\pi}^\+\cdot s_{\pi}^\+
+
d_{\pi}^\+\cdot d_{\pi}^\+)
+
(H.c.) \; ,
\end{eqnarray}
with the strength $\omega$ assumed
to be equal for the $s_{\pi}$
and $d_{\pi}$ boson terms.
Note that the forms \eqref{eq:ham2}
and \eqref{eq:vmix} are the same as those
used in the previous IBM2-CM studies
for the Hg isotopes \cite{nomura2013hg},
except that the rotational term
$\hat L\cdot\hat L$ and
the three-body boson terms
are not included in the
present work.
More details about the Hamiltonian
are found in \cite{nomura2013hg}.

The IBM2-CM Hamiltonian
is determined by associating it with
the corresponding SCMF PES \cite{nomura2012sc}.
The geometry of the IBM2-CM Hamiltonian
is realized as a $2\times2$ matrix
\begin{eqnarray}
\label{eq:pes}
 {\mathcal E}(\beta,\gamma)
=
\left(
\begin{array}{cc}
 E_{11}(\beta,\gamma) & \Omega_{13}(\beta,\gamma) \\
 \Omega_{31}(\beta,\gamma) & E_{33}(\beta,\gamma) \\
\end{array}
\right) \; ,
\end{eqnarray}
calculated
as the expectation value of the Hamiltonian
\eqref{eq:ham1}
in the boson coherent state
expressed as a direct sum
of those for the two independent
boson spaces
\cite{frank2004}.
The diagonal elements $E_{ii}$ in
\eqref{eq:pes} are expectation
values of the unperturbed
Hamiltonians, while the nondiagonal
parts $\Omega_{13}=\Omega_{31}$
are those of the mixing term \eqref{eq:vmix}.
As is conventional, the lower-energy
eigenvalue of the coherent-state matrix
\eqref{eq:pes} is taken as the
IBM2-CM energy surface.
Explicit analytic forms of
$E_{ii}$ and $\Omega_{13}$
are found in Ref.~\cite{nomura2013hg}.

The parameters for
the unperturbed Hamiltonian $\hat H_{1}$
are fixed \cite{nomura2008} so that
the expectation value of the
unperturbed Hamiltonian, that is,
the diagonal element of the coherent-state
matrix $E_{11}$, should reproduce
basic features of the RHB-SCMF PES
in the vicinity of the oblate global
minimum, with the
corresponding $\beta$ deformation
denoted $\beta_{gl}$.
The Hamiltonian for the intruder
configuration $\hat H_{3}$ is
determined in a similar way,
i.e., by associating $E_{33}$ to the
local prolate minimum
in the RHB-SCMF PES
with a larger $\beta$ deformation
denoted $\beta_{lo}$.

Having fixed the unperturbed Hamiltonian
parameters,
the offset energy $\Delta$
is extracted so that the
energy difference between
the global and local
minima in the SCMF PES should be
reproduced, that is,
the relation
\begin{eqnarray}
\label{eq:delta1}
 \left[
E_{33}(\beta_{lo},\gamma_{lo})+\Delta
\right]
- E_{11}(\beta_{gl},\gamma_{gl})
= \delta E
\end{eqnarray}
is made to hold.
$E_{11}(\beta_{gl},\gamma_{gl})$
and $E_{33}(\beta_{lo},\gamma_{lo})$
denote the diagonal elements of the
matrix \eqref{eq:pes}
calculated at the configurations
$(\beta_{gl},\gamma_{gl})$
and
$(\beta_{lo},\gamma_{lo})$.
$\delta E$ is the
energy at the prolate local minimum
relative to that at
the oblate global minimum
in the SCMF PES.
Finally,
the mixing interaction $\hat V_{\rm mix}$
is introduced, and the coherent-state
matrix \eqref{eq:pes} is diagonalized.
The mixing strength
$\omega$ is determined
so that the IBM2-CM PES, i.e., the lowest
eigenvalue of the matrix, should reasonably
reproduce the
$\gamma$ dependence of
that part of the PES
between the
oblate and prolate minima.

A special treatment should be
made of the offset energy $\Delta$.
This parameter was originally
defined \cite{duval1982}
as a constant added to the
energy eigenvalue for the ground state
$E_{3}(0^+_{gs})$ of the
unperturbed intruder Hamiltonian
so that it should become higher than
the ground-state energy
eigenvalue
$E_{1}(0^+_{gs})$ of the
unperturbed normal configuration
by an amount $\delta E'$,
which is approximately
set equal to the observed $0^+_2$
excitation energy:
\begin{eqnarray}
\label{eq:delta1a}
 E_{3}(0^+_{gs})+\Delta
=E_{1}(0^+_{gs})+\delta E' \; .
\end{eqnarray}
In the previous mapped IBM2-CM
study of the shape coexistence in
Pb isotopes \cite{nomura2012sc},
instead of the $\Delta$ values determined
by the formula \eqref{eq:delta1},
those $\Delta$ values obtained
from the formula \eqref{eq:delta1a},
but with $\delta E'$ replaced
with $\delta E$, i.e.,
\begin{eqnarray}
\label{eq:delta2}
 E_{3}(0^+_{gs})+\Delta
=E_{1}(0^+_{gs})+\delta E
\end{eqnarray}
were used to
compute spectroscopic properties.
In other words,
a different $\Delta$ value
from that obtained from
the PES-mapping procedure
was employed when diagonalizing
the Hamiltonian.
The $0^+_{gs}$ energy eigenvalue
of the (unperturbed) boson
Hamiltonian comprises the energies
gained by deformations
at the mean-field level,
and quantum correlation
energies that are beyond
the mean-field level.
While among those parameters involved
in the IBM2-CM Hamiltonian
the $\Delta$ parameter should
be most relevant to account
for the ground-state
correlation effects,
the $\Delta$ values in
\eqref{eq:delta1} 
are defined in terms only of
the deformation energies at
the mean-field level.
Since the quantum correlation
energies are included
only through the diagonalization
of the boson Hamiltonian
in the laboratory frame,
it appears to be
more appropriate to employ
the $\Delta$ values obtained
by \eqref{eq:delta2}
for calculating spectroscopic properties
than those from \eqref{eq:delta1}.

In fact, it was shown
in \cite{nomura2012sc} that
the $\Delta$ values obtained from
the formula \eqref{eq:delta1}
turned out to
be considerably smaller than
those from \eqref{eq:delta2},
and that the deformed
intruder $0^+$ state
became the ground state,
which contradicts
the empirical finding.
The reason for this was
that for the Pb isotopes the normal
configuration was associated with
the spherical mean-field minimum
and did not gain energies arising from
deformations.
In the present case,
the $\Delta$ values from \eqref{eq:delta1}
are shown to be systematically
larger than those from \eqref{eq:delta2}
approximately by 0.6 MeV.
This is because
the normal configuration gains
substantial amounts of
correlation energies
through the diagonalization
due to large quadrupole-quadrupole
interaction.
Once the configuration mixing
is carried out, the resultant
energy levels for nonyrast states
are overall substantially higher
than the experimental ones
and, in particular, the observed orders
of levels are not reproduced.
Another notable consequence
of using the $\Delta$ values from
\eqref{eq:delta1} is that
the $E2$ transitions of the
nonyrast states, particularly
the $0^+_2$ state, turn out to
be negligibly small.
This was also encountered
in previous applications
of the IBM2-CM mapping
(e.g., \cite{nomura2016zr,nomura2018cd}),
in which
the $\Delta$ values calculated by the formula
\eqref{eq:delta1} were used for
spectroscopic calculations,
and $E2$ transitions for nonyrast
states in some nuclei
were not properly reproduced.

The $\Delta$ values of
\eqref{eq:delta2}, however,
do not necessarily reproduce
accurately the features
of the SCMF PESs,
in particular, the oblate-prolate
energy balance,
that is, difference
in energy between the
oblate and prolate minima.
The uncertainty of the parameters
for the IBM2-CM was also noticed
in Ref.~\cite{frank2004},
in which the IBM2-CM parameters
that gave a good reproduction of
low-lying states in the $^{186}$Pb
nucleus were shown to
produce only two mean-field
minima in the energy surface
after configuration mixing.
This is in contrast with
the empirical understandings
\cite{andre00,ojala2022-186Pb}
that the structure of $^{186}$Pb
is characterized by the
coexistence of the spherical,
oblate, and prolate shapes.
It seems to be an open
question to develop a consistent
way of deriving $\Delta$ values
that simultaneously give
reasonable descriptions of
the PESs and spectroscopic properties
in the IBM2-CM mapping.

The IBM2-CM Hamiltonian with the
parameters obtained by the aforementioned
procedure is diagonalized in the
space \eqref{eq:space} using
the boson $m$-scheme
basis \cite{nomura2012sc}.
As noted, the IBM2-CM calculations
are performed specifically
for $^{114-122}$Te, for which
the self-consistent
calculations suggest two minima
\cite{suzuki2024}.
The standard
EDF-mapped IBM-2 calculations,
i.e., those that consist only
of a single (normal) boson configuration
and do not include
the configuration mixing,
are carried out for
the other isotopes $^{108-112}$Te
and $^{124,126}$Te.
In order to compare with
the IBM2-CM results,
the standard IBM-2 mapping
calculations are also made for $^{114-122}$Te.
Furthermore,
the standard IBM-2 with a
single configuration is below
referred to as IBM-2,
for the sake of simplicity.

%
%
\begin{figure}
\begin{center}
\includegraphics[width=\linewidth]{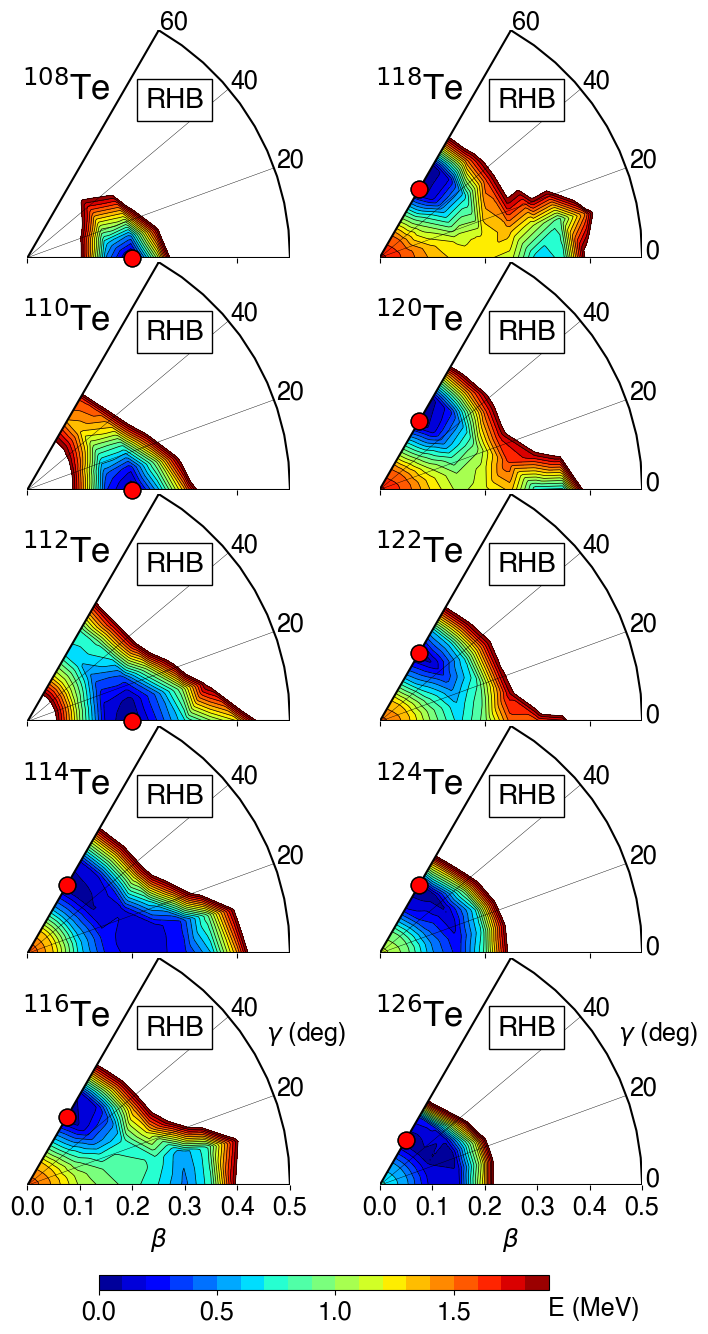}
\caption{
Potential energy surfaces
in terms of the quadrupole
deformations $\beta$ and $\gamma$ for the
$^{108-126}$Te isotopes calculated
by the constrained
RHB-SCMF method employing
the DD-PC1 EDF and a separable pairing force.
The global minimum is indicated
by the solid circle.
See the main text for details.
}
\label{fig:pes1}
\end{center}
\end{figure}

%
%
\begin{figure}
\begin{center}
\includegraphics[width=\linewidth]{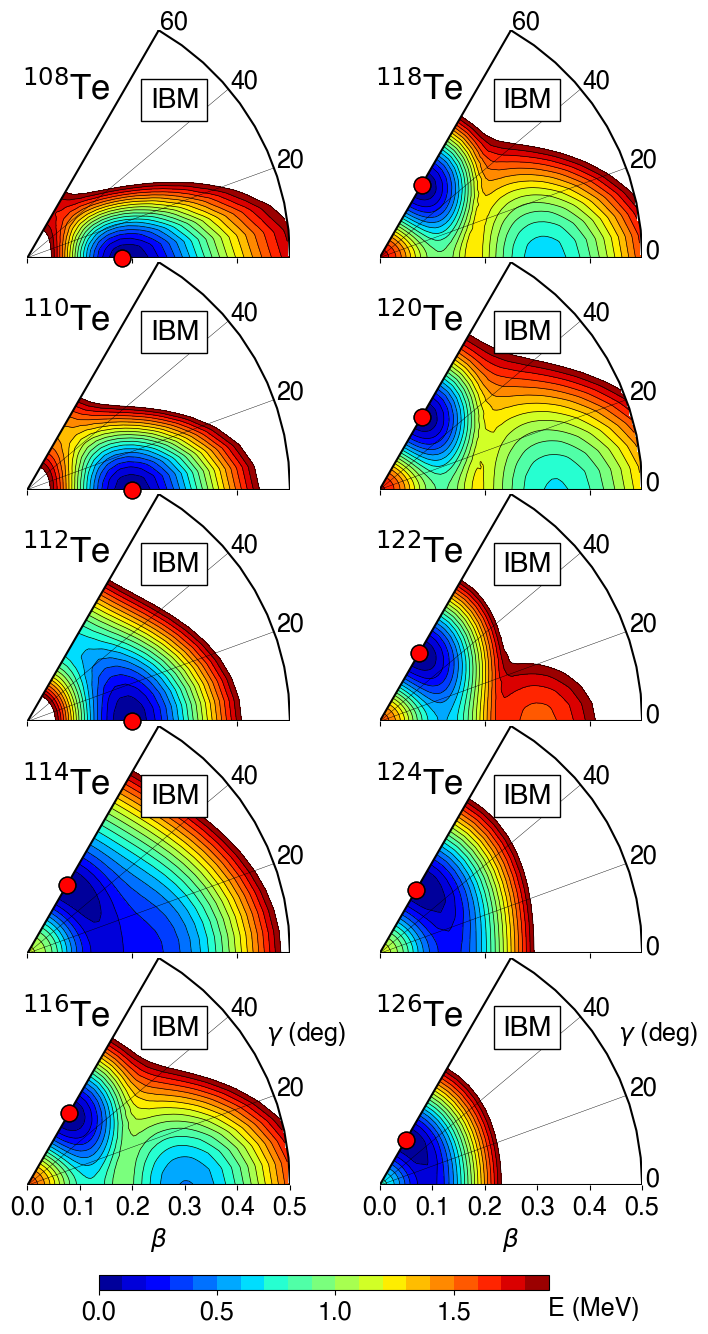}
\caption{
Same as the caption to Fig.~\ref{fig:pes1},
but for the mapped bosonic PESs.
}
\label{fig:pes2}
\end{center}
\end{figure}

%
%
\begin{figure}[ht]
\begin{center}
\includegraphics[width=\linewidth]{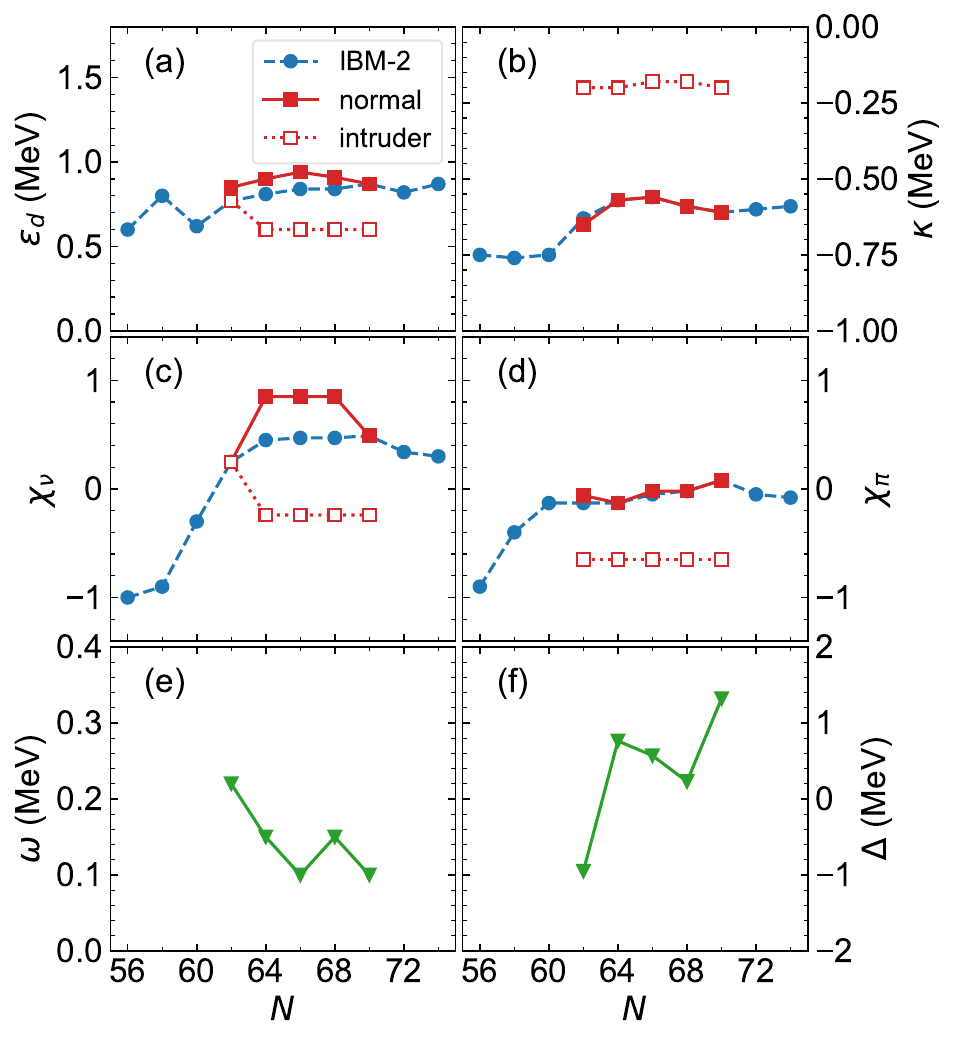}
\caption{
Derived Hamiltonian parameters for the
IBM-2 and those for the IBM2-CM
corresponding to
the normal and intruder configurations,
mixing strength and energy offset
$\Delta$ of \eqref{eq:delta2}
for the studied Te nuclei.
}
\label{fig:para}
\end{center}
\end{figure}

%
%
\begin{figure*}[ht]
\begin{center}
\includegraphics[width=.8\linewidth]{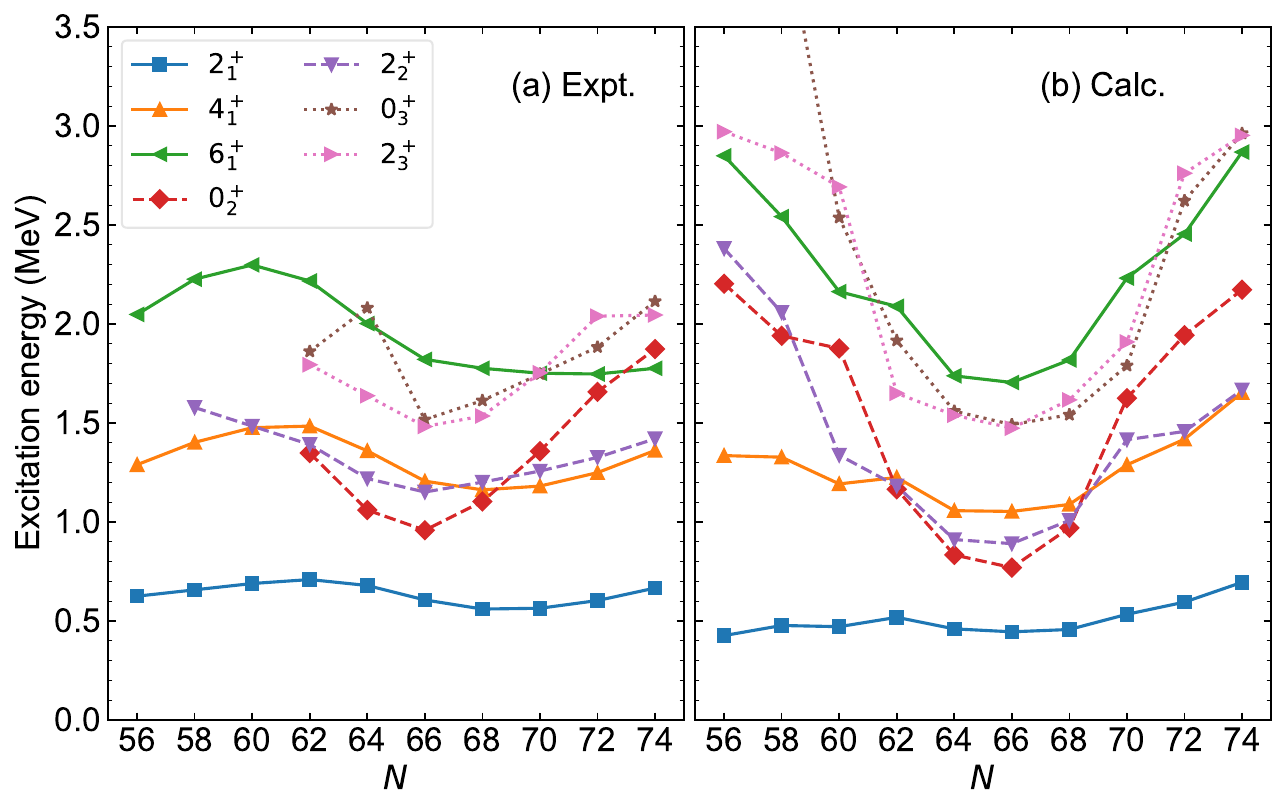}
\caption{
(a) Experimental
\cite{data,mihai2011,doncel2017,vonspee2024}
and (b) calculated
excitation energies of low-lying
states for the even-even
$^{108-126}$Te isotopes.
The calculations are based on
the IBM2-CM for $^{114-122}$Te,
and on the IBM-2
with a single configuration
for other nuclei.
}
\label{fig:level}
\end{center}
\end{figure*}

%
%
\begin{figure}[ht]
\begin{center}
\includegraphics[width=.8\linewidth]{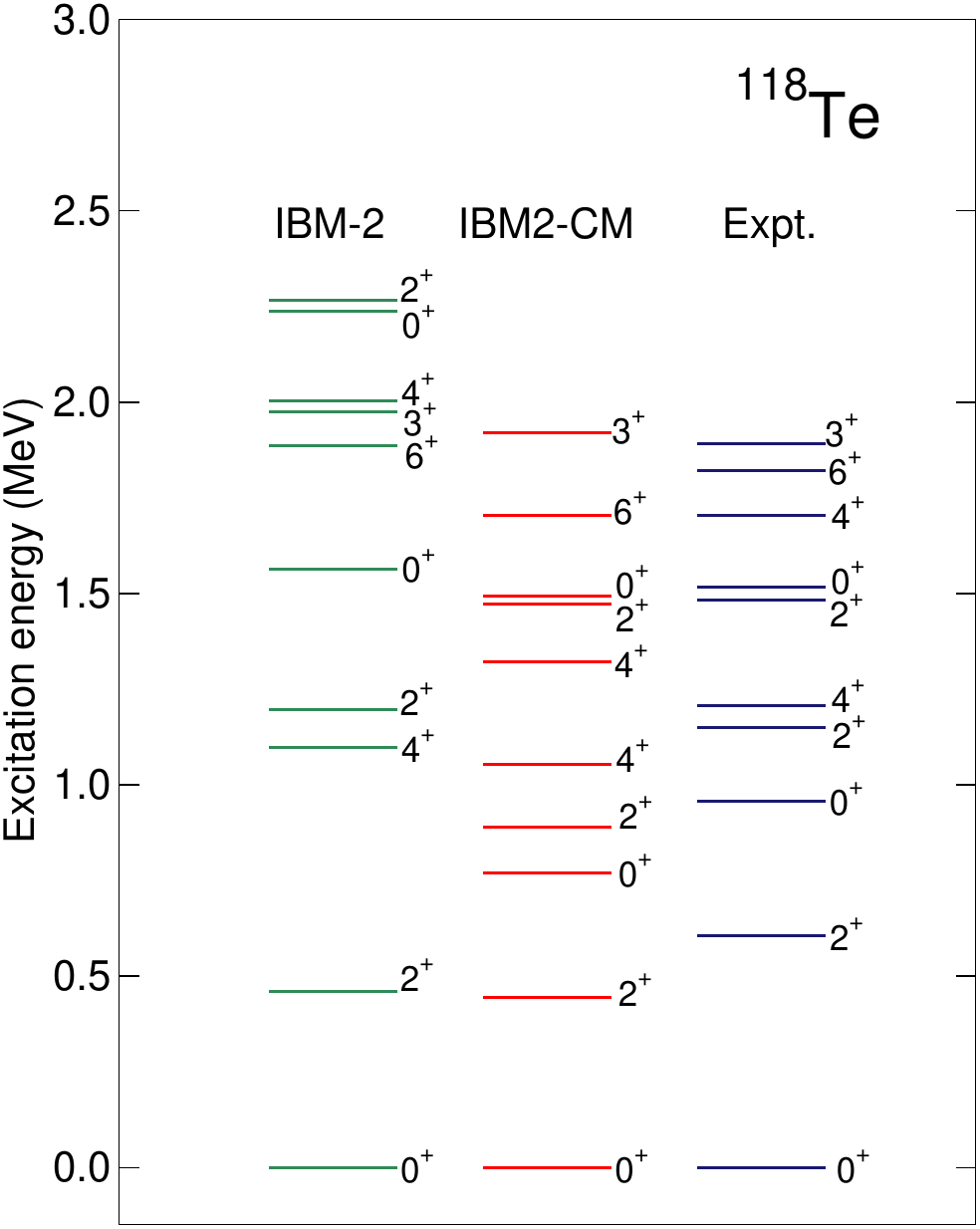}
\caption{
Predicted low-energy spectra
for $^{118}$Te from the IBM-2
and IBM2-CM, and the experimental
data \cite{mihai2011}.
}
\label{fig:te118}
\end{center}
\end{figure}

\section{Results\label{sec:results}}

\subsection{Energy surfaces}

Figures~\ref{fig:pes1}
and \ref{fig:pes2}
show, respectively, the RHB-SCMF
PESs and corresponding bosonic
PESs for the nuclei $^{108-126}$Te.
For obtaining the
IBM2-CM PESs for $^{114-122}$Te
the $\Delta$ values
of the formula \eqref{eq:delta1} are used.
The RHB-SCMF PESs suggest
a transition from prolate to oblate
shapes along the isotopic chain
from near the neutron major
shell $N=50$ toward
the other end, $N=82$.
There appears pronounced
coexistence of the oblate
global minimum in
$\beta\approx 0.15-0.2$
and a prolate or triaxial minimum
at $\beta\approx 0.3$
in transitional
nuclei $^{114-120}$Te.
The prolate local minimum is
also visible in $^{122}$Te,
but is much more minor than
in $^{114-120}$Te.
More detailed
discussions about the shape evolution
inferred from the
mean-field energy surfaces
are found in Ref.~\cite{suzuki2024}.
One can see in Fig.~\ref{fig:pes2}
that basic properties and systematic
of the mapped bosonic PESs,
such as the location of the minima,
softness in the $\beta$
and $\gamma$ deformations,
depth of the potential valley,
etc., resemble those
of the original RHB-SCMF PESs.
A noticeable difference
is that the bosonic
PESs are rather flat as compared with
the SCMF ones.
The reason for this is that the
IBM space is essentially
comprised
of valence nucleons only,
while the SCMF model is
generally built on
an entire fermion model space
consisting of a larger
number of single-particle
states.

In addition,
in view of the recent finding
that multiple shapes emerge in midshell
Cd isotopes \cite{garrett2019},
using two-boson configurations
for the midshell Te isotopes
may seem to be insufficient
to fully take into account
possible effects of
mixing of shapes other than
the oblate and prolate ones
on the low-energy structure.
For the SCMF-PES mapping procedure,
however, especially
if the potential is soft, such as
in the case of $^{114}$Te, constraining
the parameters of more than
two unperturbed IBM-2 Hamiltonians,
energy offsets and
mixing strengths would be practically
difficult, and this is the reason
why the present calculation
is restricted to two-boson spaces.
The prolate-oblate
energy balance of the
mean-field minima
and $\gamma$ softness 
are also rather sensitive to the
EDFs used in the SCMF calculations.
This point will be discussed
in more detail in Sec.~\ref{sec:edf}.

The parameters for the IBM2-CM
Hamiltonian derived from the mapping
are given in Fig.~\ref{fig:para}.
In most cases,
the derived parameters for both the
unperturbed normal
and intruder configurations
show gradual variation as functions of $N$.
Rather significant changes in
the parameters $\chi_{\nu}$ and
$\chi_{\pi}$ from $N=58-62$ reflect
the transition from the
prolate to
$\gamma$-soft shapes, as suggested
at the SCMF level.
A particularly large mixing
strength $\omega$ for $^{114}$Te
reflects the pronounced
$\gamma$ softness in the PES.
A negative $\Delta$ value for
$^{114}$Te is also of interest.
The reason for this is that
the normal oblate configuration
gains large deformation energies
due to the strong
quadrupole-quadrupole interaction
as compared with the
intruder configuration
[see \eqref{eq:delta2}],
and that the prolate local minimum
is very close in energy
to the oblate global minimum
(Fig.~\ref{fig:pes1}).

\subsection{Low-energy spectra}

Figure~\ref{fig:level}
depicts systematic of the
excitation energies of
low-lying states for $^{108-126}$Te.
The experimental
$2^+_1$ energy level,
shown in Fig.~\ref{fig:level}(a),
exhibits gradual evolution
with $N$ and becomes lowest at $N=68$.
The observed $4^+_1$ and $6^+_1$
energy levels exhibit
slightly different patterns from
that of the $2^+_1$ level,
that is,
the $4^+_1$ energy level
exhibits a weak parabolic dependence
on $N$, and the $6^+_1$ level
stays rather constant from $N=68-74$.
Experimentally,
the $2^+_1$, $4^+_1$, and $6^+_1$
levels exhibit inverse parabolas
with maximal values
at $N=62$ (for the
$2^+_1$ and $4^+_1$ states)
and $N=60$ (for the $6^+_1$ state).
The observed levels
for the nonyrast states
$0^+_{2,3}$ and $2^+_{2,3}$
exhibit a parabolic behavior,
and become lowest in energy at
the neutron midshell $N=66$.
The $0^+_2$ and $2^+_3$ energy
levels are experimentally suggested
to vary especially sharply.

Figure~\ref{fig:level}(b) shows
the predicted excitations energies,
obtained from the IBM2-CM for
$^{114-122}$Te and
from the IBM-2 with a single configuration
for the other Te nuclei.
The IBM2-CM results
for all spectroscopic properties
discussed in the following
are those obtained
using the $\Delta$
values of \eqref{eq:delta2}.
The overall systematic of the calculated
$2^+_1$ levels is rather consistent with
the pattern observed experimentally.
The calculated $4^+_1$
energy levels exhibit a weak
parabolic dependence on $N$.
The calculated ratios
$R_{4/2}=E(4^+_1)/E(2^+_1)$
of the $4^+_1$ to $2^+_1$
excitation energies take
almost constant values
$R_{4/2}\approx2.3-2.4$
for $^{114-126}$Te,
which are slightly larger than
the experimental values
$R_{4/2}\approx2.0$.
For the $^{108,110}$Te nuclei,
the IBM-2 calculation
with a single
(prolate) configuration yields
a more rotational-like spectrum
with the $R_{4/2}$ ratios,
3.13 and 2.78, respectively,
which are higher than the
experimental values
$R_{4/2}=2.06$ and 2.13.
The differences are due to the
fact that the RHB-SCMF
PES exhibits a very sharp potential
valley, and to reproduce this feature
both the $\chi_{\nu}$ and $\chi_{\pi}$
parameters have to be large
in magnitude, being close to the
SU(3) limit,
$-\sqrt{7}/2$ \cite{IBM},
and the resultant IBM-2 Hamiltonian
produces a rotational-like spectrum.
Furthermore,
the IBM-2 space for a single
configuration for these
nuclei may be too restricted to
describe reasonably the energy levels
with spin $I\geqslant6$.
On the neutron-rich side with
$N\geqslant70$,
the observed $4^+_1$ systematic
is more or less reproduced in the
present calculation.

The $6^+_1$ energy
levels predicted
in the IBM2-CM exhibit a stronger
parabolic dependence on $N$ than
the $4^+_1$ ones,
but experimentally
the $4^+_1$ and $6^+_1$
energy levels are
quite similar in systematic.
The quantitative differences
between the
$4^+_1$ and $6^+_1$
energy levels within the theory
could be attributed
to the facts that,
as mentioned above,
the boson model space
becomes more restricted
when approaching the neutron
major shells $N=50$ and 82,
pushing up
higher-spin states
with $I\geqslant 6$,
and that the mixing in these
low-lying states is so strong
that significant
degrees of level
repulsion occur.
The particularly
low-lying experimental
$6^+_1$ energy level
with respect to the $4^+_1$ one
on the neutron-rich side does
not look like a typical
collective band structure.
To account for this energy-level
pattern, some additional correlations,
e.g., couplings to
noncollective degrees of freedom,
may need to be taken into
account in the
IBM framework.

The predicted nonyrast levels
display a parabolic pattern
and become particularly low in energy
in the isotopes $^{114-120}$Te,
with minimal energies at $N=66$.
This systematic resembles that
of the experimental data and is
considered to be a signature
of the prolate-oblate shape coexistence.

Concerning the nuclei $^{114-122}$Te,
the IBM-2 calculations using
only a single oblate configuration
produce the excitation energies
for the yrast states that
essentially do not differ
from those within the IBM2-CM.
Notable differences between
the IBM2-CM and IBM-2
are found in that
the IBM-2 generally produces
the nonyrast levels
that are much higher than those
from the IBM2-CM.
As an illustrative example,
Fig.~\ref{fig:te118} shows
low-energy spectra for $^{118}$Te
computed by the IBM2-CM and IBM-2.
The IBM2-CM predicts low-lying
$0^+_2$ and $0^+_3$ energy levels
that are more consistent with
experiment than the IBM-2.
An improvement over the IBM-2
is that not only the excited $0^+$
energy levels, but also
the order of levels,
except for the $4^+_2$ level,
appears to be reproduced
better in the IBM2-CM.

Note also that the present IBM2-CM and IBM-2
results for the low-energy levels
are qualitatively similar to those
from the RHB+QCH calculations
in Ref.~\cite{suzuki2024},
particularly the one that
employed the default value
of the pairing
strengths constant for all isotopes.
Some quantitative differences
from the present results
are that the QCH predicted the
$2^+_1$, $4^+_1$, and
$0^+_2$ energy levels
to be lowest at $^{116}$Te,
and produced a local
inverse parabolic pattern
of the $2^+_1$ and $4^+_1$
levels with maximal energies
at $N=58$ or 60.

To see the importance of the configuration
mixing, Table~\ref{tab:wf} gives
overlap probabilities (in percent) of
the basis states
corresponding to the intruder boson
space of $N_{\pi}=3$
and the eigenfunctions of the IBM2-CM for
low-spin states.
From the table one can see that,
in most of the states, significant
degrees of intruder prolate
contributions are present even
in the yrast states.
Particularly strong mixing is
suggested for the low-lying states
in $^{114}$Te and $^{116}$Te.
For all those nuclei shown in
the table, the $0^+_2$ state
mainly comes from the
prolate intruder configuration.
For $^{122}$Te, the mixing is
minor, and the intruder
configuration does not seem to
play an important role,
except for the $0^+_2$ 
and $2^+_3$ wave functions.

%
\begin{table}
\caption{\label{tab:wf}
Contributions (in percent)
of the intruder configurations
in the IBM2-CM wave functions for
low-spin states of $^{114-122}$Te.
}
 \begin{center}
 \begin{ruledtabular}
  \begin{tabular}{lccccc}
& $^{114}$Te & $^{116}$Te & $^{118}$Te & $^{120}$Te & $^{122}$Te \\
\hline
$0^+_1$ & $40$ & $21$ & $9$ & $15$ & $3$ \\
$0^+_2$ & $63$ & $81$ & $91$ & $86$ & $96$ \\
$0^+_3$ & $46$ & $11$ & $6$ & $11$ & $3$ \\
$2^+_1$ & $48$ & $30$ & $12$ & $18$ & $3$ \\
$2^+_2$ & $56$ & $64$ & $82$ & $70$ & $14$ \\
$2^+_3$ & $55$ & $34$ & $22$ & $36$ & $87$ \\
$4^+_1$ & $57$ & $54$ & $33$ & $31$ & $4$ \\
$6^+_1$ & $69$ & $78$ & $79$ & $62$ & $7$ \\
 \end{tabular}
 \end{ruledtabular}
 \end{center}
\end{table}

%
%
\begin{figure}[ht]
\begin{center}
\includegraphics[width=.8\linewidth]{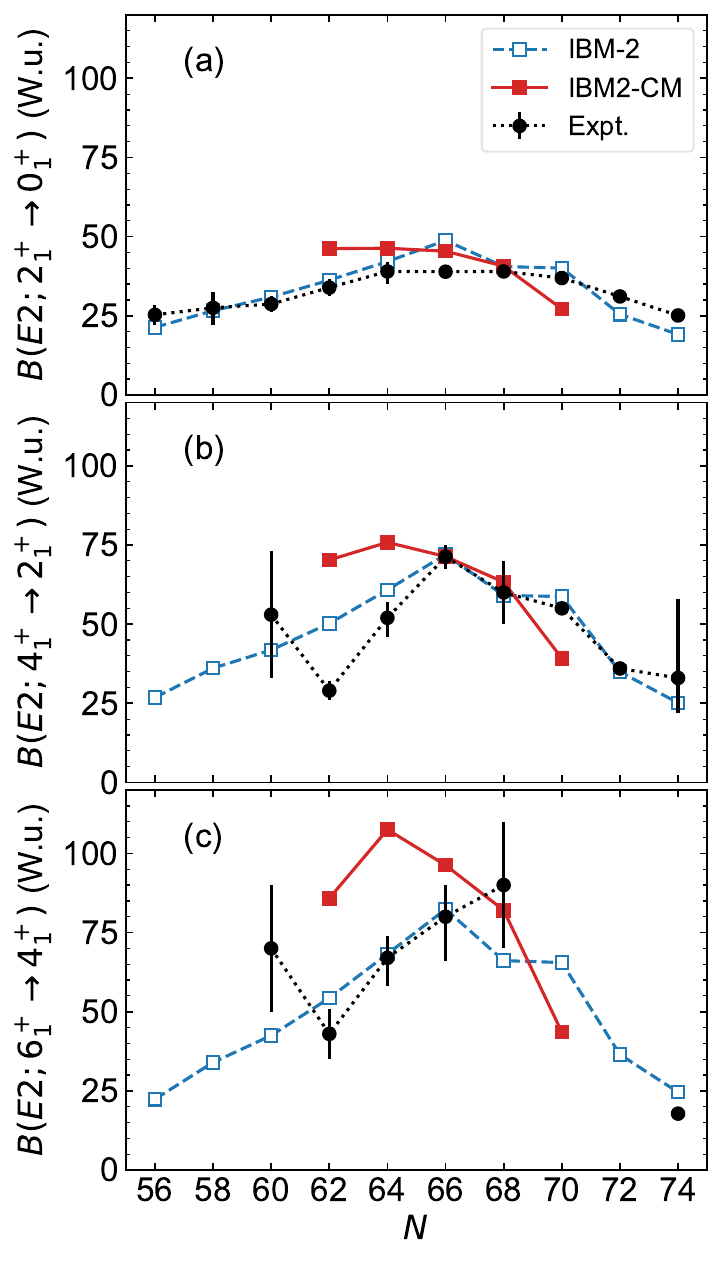}
\caption{
(a) $\be 2 1 0 1$,
(b) $\be 4 1 2 1$,
and (c) $\be 6 1 4 1$ values
in Weisskopf units (W.u.)
for Te isotopes calculated with the
IBM2-CM and IBM-2.
Experimental data are adopted from
Refs.~\cite{doncel2017,testov2021,
pascu2025,cederlof2023-118Te,mihai2011,
saxena2014,PRITYCHENKO2016,data}.
See the main text for details.
}
\label{fig:be2-y}
\end{center}
\end{figure}

%
%
\begin{figure*}[ht]
\begin{center}
\includegraphics[width=.7\linewidth]{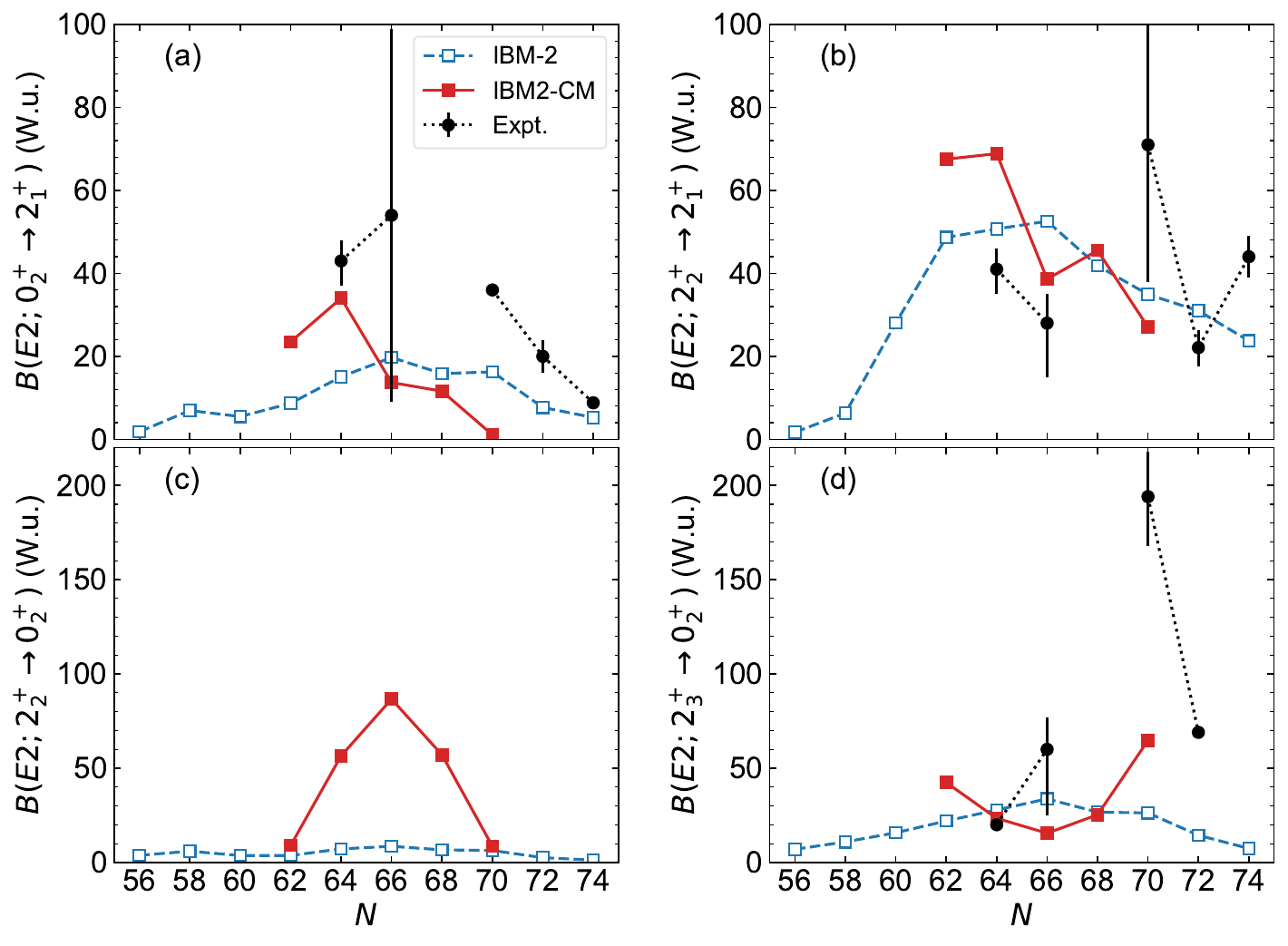}
\caption{
Same as the caption to
Fig.~\ref{fig:be2-y},
but for the
(a) $\be 0 2 2 1$,
(b) $\be 2 2 2 1$,
(c) $\be 2 2 0 2$, and
(d) $\be 2 3 0 2$
values.
Experimental data are taken from
Refs.~\cite{vonspee2024,mihai2011,
hicks2005-122Te,hicks2017-124Te}.
The experimental $\be 2 3 0 2$
value for $^{116}$Te
is a lower limit, $>20$ W.u.
\cite{vonspee2024}.
Also the experimental
$\be 0 2 2 1$ value for $^{122}$Te
\cite{hicks2005-122Te}, and
$\be 2 2 2 1$ and $\be 2 3 0 2$
values for $^{124}$Te
\cite{hicks2017-124Te} are
upper limits.
}
\label{fig:be2-ny}
\end{center}
\end{figure*}

%
%
\begin{figure}[ht]
\begin{center}
\includegraphics[width=.8\linewidth]{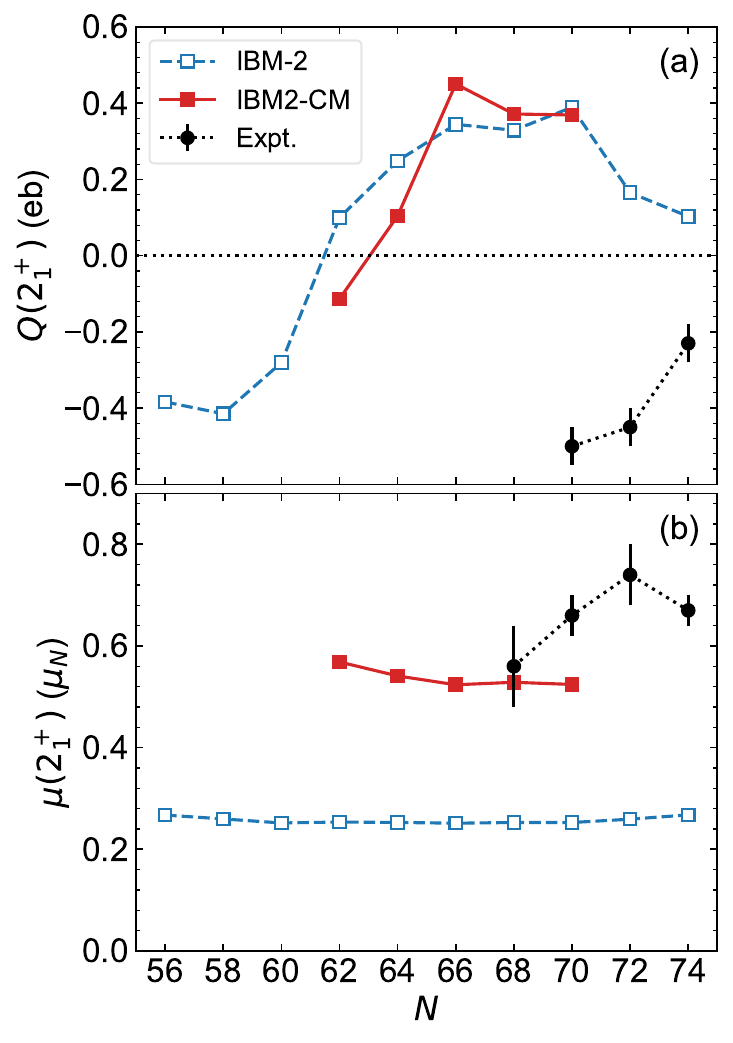}
\caption{
(a) Spectroscopic electric quadrupole
moments and (b) magnetic dipole moments
for the $2^+_1$ state for the Te isotopes.
The calculated results from the IBM-2
and IBM2-CM are shown.
Experimental data are taken from
ENSDF database \cite{data}.
}
\label{fig:mom}
\end{center}
\end{figure}

\subsection{Electromagnetic properties}

Using the wave functions
resulting
from the mapped IBM2-CM Hamiltonian,
transition properties for
given states are computed.
In the IBM2-CM, the $E2$ operator is
given as
\begin{eqnarray}
 \hat T^{(E2)}=\sum_{i=1,3}
\hat P_{i}
(e_{\nu,i} \hat Q_{\nu,i}
+e_{\pi,i} \hat Q_{\pi,i})
\hat P_{i} \; ,
\end{eqnarray}
where $e_{\rho,i}$ and $\hat Q_{\rho,i}$
are neutron or proton
effective charges and
quadrupole operators \eqref{eq:ham2}
for the
normal ($i=1$) and intruder ($i=3$)
configurations, respectively.
The effective charges,
$e_{\nu,1}=e_{\pi,1}=0.085$ $e$b and
$e_{\nu,3}=e_{\pi,3}=0.110$ $e$b,
which are similar to those used
for Cd isotopes \cite{nomura2018cd},
are adopted in the IBM2-CM.
For the IBM-2,
the boson charges
$e_{\nu}=e_{\pi}=0.095$ $e$b
are employed so as to
reproduce the overall systematic
with $N$ of the experimental
$\be 2 1 0 1$,
$\be 4 1 2 1$, and
$\be 6 1 4 1$ values.
The $M1$ operator in the IBM2-CM reads
\begin{eqnarray}
 \hat T^{(M1)}=
\sqrt{\frac{3}{4\pi}}
\sum_{i=1,3}
\hat P_{i}
(g_{\nu,i} \hat L_{\nu}
+g_{\pi,i} \hat L_{\pi})
\hat P_{i} \; ,
\end{eqnarray}
where
$\hat L_{\rho}
=\sqrt{10}[d^{\+}_{\rho}\times\tilde d_{\rho}]^{(1)}$
is the boson angular-momentum
operator. For the
effective $g$ factors
$g_{\rho,i}$, the empirical values
obtained in Ref.~\cite{rikovska1989}
are adopted:
$g_{\nu,1}=g_{\nu,3}=0.24$ $\mu_N$
(nuclear magneton) and
$g_{\pi,1}=g_{\pi,3}=0.5$ $\mu_N$.

Figure~\ref{fig:be2-y} exhibits
calculated
$\be 2 1 0 1$, $\be 4 1 2 1$, and
$\be 6 1 4 1$ values in Weisskopf units
(W.u.) for the Te isotopes
in the IBM2-CM and IBM-2.
The $\be 2 1 0 1$
data are adopted from
Ref.~\cite{testov2021} for $^{110}$Te,
Ref.~\cite{pascu2025} for $^{116}$Te,
Ref.~\cite{cederlof2023-118Te} for $^{118}$Te,
and Ref.~\cite{PRITYCHENKO2016}
for all the other Te isotopes.
The data for the
$\be 4 1 2 1$ and $\be 6 1 4 1$
values are taken from
Ref.~\cite{doncel2017} for
$^{112,114,120,122,124}$Te,
Ref.~\cite{pascu2025} for $^{116}$Te,
and Ref.~\cite{data} for $^{126}$Te.
The experimental
$\be 4 1 2 1$ and $\be 6 1 4 1$
values for $^{118}$Te
are adopted from
Ref.~\cite{cederlof2023-118Te}
and Ref.~\cite{mihai2011},
respectively.

From Fig.~\ref{fig:be2-y}(a) one can
observe gradual evolution of the
experimental $\be 2 1 0 1$ values.
The IBM-2 reproduces this trend
and gives the maximal $\be 2 1 0 1$
value at the middle of the shell $N=66$.
In the IBM2-CM,
the maximal $\be 2 1 0 1$ value,
46.3 W.u., is obtained rather
at $N=64$, which is slightly
larger than the recent experimental value
39$^{+4}_{-3}$ W.u. \cite{pascu2025}
The IBM2-CM provides a particularly
large $\be 2 1 0 1$ value, 46.2 W.u.,
for $^{114}$Te, which overestimates
the experimental value \cite{data}.
The $\bej 4 1 2 1$ transition rates
are experimentally suggested to be
maximal at the midshell $N=66$.
This behavior is reasonably
reproduced by the IBM-2.
The IBM2-CM provides for $^{118}$Te
and $^{120}$Te the $\be 4 1 2 1$
values that are consistent
with the corresponding
experimental values,
71.3$\pm9.8$ W.u.
\cite{cederlof2023-118Te} and
60$\pm10$ W.u. \cite{saxena2014},
respectively.

What is worth noticing is a
reduction of the observed
$\be 4 1 2 1$ value at $^{114}$Te,
being of the same order
of magnitude as the $\be 2 1 0 1$ value.
This local pattern
is not reproduced either by
the IBM2-CM or IBM-2.
Earlier IBM2-CM \cite{rikovska1989}
and large-scale shell model (LSSM)
\cite{kaneko2021} calculations
also overestimated
the $\be 4 1 2 1$ value for $^{114}$Te.
The observed ratio
$B_{4/2}=\be 4 1 2 1/\be 2 1 0 1<1$
in $^{114}$Te indicates a deviation
from the interpretation
in the collective model, in which
$B_{4/2}>1$ is expected.
This feature resembles that
found in heavier nuclei, e.g., $^{168}$Os
\cite{grahn2016},
which is often referred to
as the $B(E2)$ anomaly.
The anomalous reduction of
$\be 4 1 2 1$ at $^{114}$Te
has been reported in
Refs.~\cite{cakirli2004,moeller2005-te114},
in which IBM calculations
were performed assuming
the pure vibrational U(5)
limit of the Hamiltonian.
These IBM calculations
reproduced energy spectra,
but were unable to explain
the $B(E2)$ systematic in question.
The $B(E2)$ anomaly has been
studied more recently with
the IBM \cite{wang2023,zhang2025,teng2025}
that considered higher-order
boson terms in the Hamiltonian.
From symmetry-based analyses
these studies indicated the
importance of triaxiality for
accounting for the anomalous $B(E2)$
systematic.
The RHB-SCMF PES calculated in
the present study for $^{114}$Te
suggests a pronounced $\gamma$ softness
(in Fig.~\ref{fig:pes1}).
It is an interesting open question
to study the influence of the
$\gamma$ softness on the
$B(E2)$ values obtained from
the IBM2-CM mapping that
includes some additional interaction
terms in the Hamiltonian,
as has been made in the aforementioned
calculations \cite{wang2023,zhang2025,teng2025}.

From Fig.~\ref{fig:be2-y}(c)
one can see that
experimentally
the behavior of the
$\be 6 1 4 1$ values
is basically similar
to that of the $\be 4 1 2 1$,
in particular,
the local pattern at $^{114}$Te.
The present IBM-2 generally reproduces
the experimental $\be 6 1 4 1$
values for $^{114,116,118}$Te.
One notices that in some cases
the IBM-2 seem to give a
better description of
the $\be 4 1 2 1$
and $\be 6 1 4 1$ values
than the IBM2-CM.
These results may indicate
that in the IBM mapping,
at least in its current
implementation,
the inclusion of
the configuration mixing
does not necessarily lead to
a complete description
of the transition properties.
For instance,
the overestimates of the
$\be 4 1 2 1$ and $\be 6 1 4 1$
values for $^{114}$Te and $^{116}$Te
could be attributed to
the fact that the considerable
amounts of the deformed
intruder configurations
are included in both
the $6^+_1$ and $4^+_1$ states
(see Table~\ref{tab:wf})
and give rise to
unexpectedly strong quadrupole
collectivity.

Figure~\ref{fig:be2-ny} shows
the reduced $E2$ transition strengths
for nonyrast states,
$\be 0 2 2 1$, $\be 2 2 2 1$,
$\be 2 2 0 2$, and $\be 2 3 0 2$.
The calculated $\be 0 2 2 1$ values
in the IBM2-CM for $^{114}$Te and $^{116}$Te
are rather large,
reflecting
the strong mixing of the oblate
and prolate shapes.
For $^{116}$Te, in particular,
the IBM2-CM value of $\be 0 2 2 1$
is of the same order of
magnitude as the
experimental value 43$^{+6}_{-5}$ W.u.
While the $\bej 0 2 2 1$ transition
is expected to be large for the
midshell nuclei in which configuration
mixing is supposed to play a
relevant role,
in both the IBM2-CM and IBM-2
this transition is suggested to
be rather weak for $^{118,120,122}$Te.
For the IBM-2, this is probably
related to
the fact that the $0^+_2$ energy levels
are predicted to be
systematically much higher
than the experimental ones
(see Fig.~\ref{fig:te118} in the
case of $^{118}$Te).
In the IBM2-CM,
the underestimates of
$\be 0 2 2 1$ for
$^{118,120,122}$Te can be
explained in terms of
the difference in structure
of the wave functions
for the $0^+_2$ and $2^+_1$ states:
the intruder configuration
dominates nearly 90\% of the $0^+_2$
wave function, but makes
a less significant,
$<20$\%, contribution
to the $2^+_1$ wave function
(see Table~\ref{tab:wf}).
One should also notice that
the experimental $\be 0 2 2 1$
value obtained for $^{118}$Te
in Ref.~\cite{mihai2011}
has a large error bar,
and that only the upper limit
is known experimentally
for $^{122}$Te
\cite{hicks2005-122Te}.

In Fig.~\ref{fig:be2-ny}(b),
the IBM-2 appears to give
a reasonable description
of the $\be 2 2 2 1$ values.
The corresponding IBM2-CM values
for $^{114}$Te and $^{116}$Te
are considerably larger than
those of the IBM-2, reflecting
enhanced configuration mixing effects
for these states.

In Fig.~\ref{fig:be2-ny}(c),
the IBM2-CM predicts enhanced
$E2$ transitions $\bej 2 2 0 2$
for the nuclei $^{116,118,120}$Te,
while the IBM-2 gives only vanishing
$\be 2 2 0 2$ values
for all the Te nuclei.
In the IBM2-CM, the level structure
shown in Fig.~\ref{fig:level}(b)
and the strong
$\bej 2 2 0 2$ transitions imply
that the $0^+_2$ state is the bandhead of
the excited $0^+$ band and that
the $2^+_2$ state is
a member of that band.

One observes in Fig.~\ref{fig:be2-ny}(d)
that the IBM2-CM gives
strong $E2$ transitions $\bej 2 3 0 2$
for $^{114}$Te and $^{122}$Te.
From this and the energy-level structure
shown in Fig.~\ref{fig:level}(b),
the IBM2-CM suggests for these nuclei
a possible $0^+$ excited band with
the bandhead $0^+_2$ and with the $2^+_3$
state being a member of the band.
Experimentally,
the 1482-keV level in $^{118}$Te
was assigned to be the
$2^+_3$ state, and
$\be 2 3 0 2=60^{+30}_{-17}$ W.u.
was extracted.
The measured $\be 2 3 0 2$ value
for
$^{122}$Te \cite{hicks2005-122Te},
$194^{+26}_{-24}$ W.u.,
is quite large.
The IBM-2 typically gives
$\be 2 3 0 2\approx5\sim25$ W.u.
For the midshell nucleus $^{118}$Te,
the IBM-2 value of $\be 2 3 0 2$
is larger than that of the
IBM2-CM and is within
the error bar of the
experimental data \cite{mihai2011}.

In the RHB+QCH calculation of
Ref.~\cite{suzuki2024},
the $\be 2 1 0 1$ values were of the
same order of magnitude as
those from the
present IBM2-CM and IBM-2,
but the QCH values of $\be 4 1 2 1$
were systematically larger than
the present results nearly by
a factor of 2.
Both the $\be 2 1 0 1$ and
$\be 4 1 2 1$ values in the QCH
exhibited increasing patterns
toward $^{116}$Te,
at which nucleus maximal values
were obtained.
The RHB+QCH suggested considerably
larger $\be 0 2 2 1$ than
the present IBM2-CM and IBM-2
calculations for all the Te isotopes
considered here, with a maximal
value, $\be 0 2 2 1>100$ W.u.,
being reached at $^{116}$Te.

Figure~\ref{fig:mom}
shows calculated spectroscopic quadrupole
moment $Q(2^+_1)$ (in $e$b) and magnetic
dipole moment $\mu(2^+_1)$ for the
$2^+_1$ states.
The IBM2-CM and IBM-2 both show
a change in sign of the $Q(2^+_1)$
values from $N=60$ to 62 and $N=62$ to 64,
reflecting the transition from
the prolate to oblate shapes.
For $^{122-126}$Te,
the calculated $Q(2^+_1)$ values
are opposite in sign to the
experimental values \cite{data}.
Earlier IBM2-CM calculations
provided negative $Q(2^+_1)$
values for most of the Te nuclei
with mass $A\geqslant 112$.
The previous RHB+QCH model
also predicted
$Q(2^+_1)<0$ for $^{116}$Te
and $^{118}$Te \cite{suzuki2024}.
The discrepancy shown in
Fig.~\ref{fig:mom}(a) appears to
arise from the fact that
the RHB-SCMF PES
exhibit an oblate minimum,
since in the IBM-2 mapping framework,
the sign of the $Q(2^+_1)$ moment
depends on whether
the SCMF energy minimum
is on the oblate or prolate side.
For the oblate deformation,
the $\chi_{\nu}$ and $\chi_{\pi}$
parameters in the quadrupole
operator are chosen to satisfy
the sum $\chi_{\nu}+\chi_{\pi}>0$,
hence the positive
$Q(2^+_1)$ values are obtained.
If one employs other relativistic
EDFs, e.g., density-dependent
meson-exchange (DD-ME2) EDF
\cite{DDME2},
the corresponding PESs for
most of the Te isotopes
exhibit an oblate global minimum
\cite{sharma2019,suzuki2024}
as in the case of the
DD-PC1 functional,
which is used here.
The Hartree-Fock-Bogoliubov
calculations \cite{delaroche2010}
with the Gogny-D1S EDF \cite{D1S},
predicted
a prolate global minimum
from $^{108}$Te up to $^{120}$Te,
but an essentially
nearly spherical shape
for $^{122-126}$Te.
The IBM2-CM values for the $\mu(2^+_1)$
moments for the midshell nuclei
$^{114-122}$Te are substantially
larger than those from the IBM-2.
The larger
$\mu(2^+_1)$ values in the
IBM2-CM than those in the IBM-2
may be considered to be
a characteristic effect
of the mixing.
However, given that
the $2^+_1$ states
in heavier Te nuclei
with $N=70$ to 74 are predicted
to be oblate in nature in
both the IBM-2 and IBM2-CM,
as opposed to the experimental
data [see Fig.~\ref{fig:mom}(a)],
the structure of the
$2^+_1$ states may not be
properly described in the
present IBM mapping calculations,
whether the configuration
mixing is considered or not.
The predicted magnetic moments
$\mu(2^+_1)$ in Fig.~\ref{fig:mom}(b)
are thus approximate.

The IBM2-CM produces a number of
lower-spin nonyrast states at low energy.
The electromagnetic transition properties
of these states are expected to
help distinguishing intruder states
from vibrational states
in nuclei in this mass region.
In recent years, extensive
experimental work has been made
on possible shape coexistence
effects on low-lying states
in the neighboring midshell
Sn isotopes:
$^{112,114}$Sn \cite{spieker2018-sn112114},
$^{116}$Sn \cite{petrache2019-sn116},
and $^{118}$Sn \cite{ortner2020-sn118}.
In these studies,
the IBM2-CM calculations
were performed
with parameters being fitted to data.
Comparisons of the IBM2-CM results
of the $B(E2)$ values for nonyrast
states revealed that the
data followed to a good extent
the $E2$ selection rules
required by dynamical symmetries
of the IBM, but this
was not the case
for some states with specific
spin, e.g., $3^+_1$ and
$4^+_{3,4,5}$ states in $^{116}$Sn
\cite{petrache2019-sn116}.
The authors of the above studies
attributed these discrepancies
to some missing elements in the
IBM-2 wave functions, indicating
some deficiency
of the IBM2-CM framework
in reproducing the detailed
$B(E2)$ systematic.

As shown in Figs.~\ref{fig:be2-y}
and \ref{fig:be2-ny},
difficulties in reproducing
the $B(E2)$ values with the
mapped IBM2-CM appear even for
low-lying states.
This is, however, not
surprising since
the IBM2-CM parameters are
here obtained from the SCMF
mapping procedure rather than
directly fitting to the
experimental data,
and since the derived parameters
as well as the
form of the Hamiltonian
are considerably different
from those that furnish
certain (dynamical)
symmetries leading to
the expected $E2$ selection rules.
In addition, significant
degree of configuration mixing is
present in the relevant wave functions
for low-lying states,
which makes it highly nontrivial
to firmly conclude
whether the states are oblate
or prolate in nature.
It seems to be, therefore,
beyond the scope of the
present work
to reproduce all details
of the $B(E2)$ systematic
and selections rules
accurately.

Electric monopole ($E0$) transitions
between $0^+$ states are
another signature for the
configuration mixing in
the studied mass region.
In the neighboring Sn isotopes,
for example, a recent measurement
\cite{wu2025}
reported a large $\rhoe 0 3 0 2$ value
for $^{120}$Sn, indicating strong
mixing between the $0^+_2$ and $0^+_3$
states.
Within the IBM2-CM, the $E0$ operator
is given as
\begin{eqnarray}
\label{eq:e0}
 \hat T^{(E0)}=\sum_{i=1,3}
\hat P_{i}
(a_{\nu,i} \hat n_{d_\nu}
+a_{\pi,i} \hat n_{d_\pi})
\hat P_{i} \; ,
\end{eqnarray}
where $\hat n_{d_\rho}$ represents
the neutron or proton $d$-boson
number operator \eqref{eq:ham2},
and $a_{\rho,i}$
are corresponding parameters.
The formula \eqref{eq:e0} should,
in principle, contain additional
terms that depend only on the
neutron and proton boson numbers
fixed for a given nucleus
for both configurations.
These additional
terms are relevant to study
intrinsic quantities like
charge radii, but do not contribute
to excited states, hence omitted
in the following
discussion, which is rather focused
on the $E0$ transitions.

%
%
\begin{figure}[ht]
\begin{center}
\includegraphics[width=\linewidth]{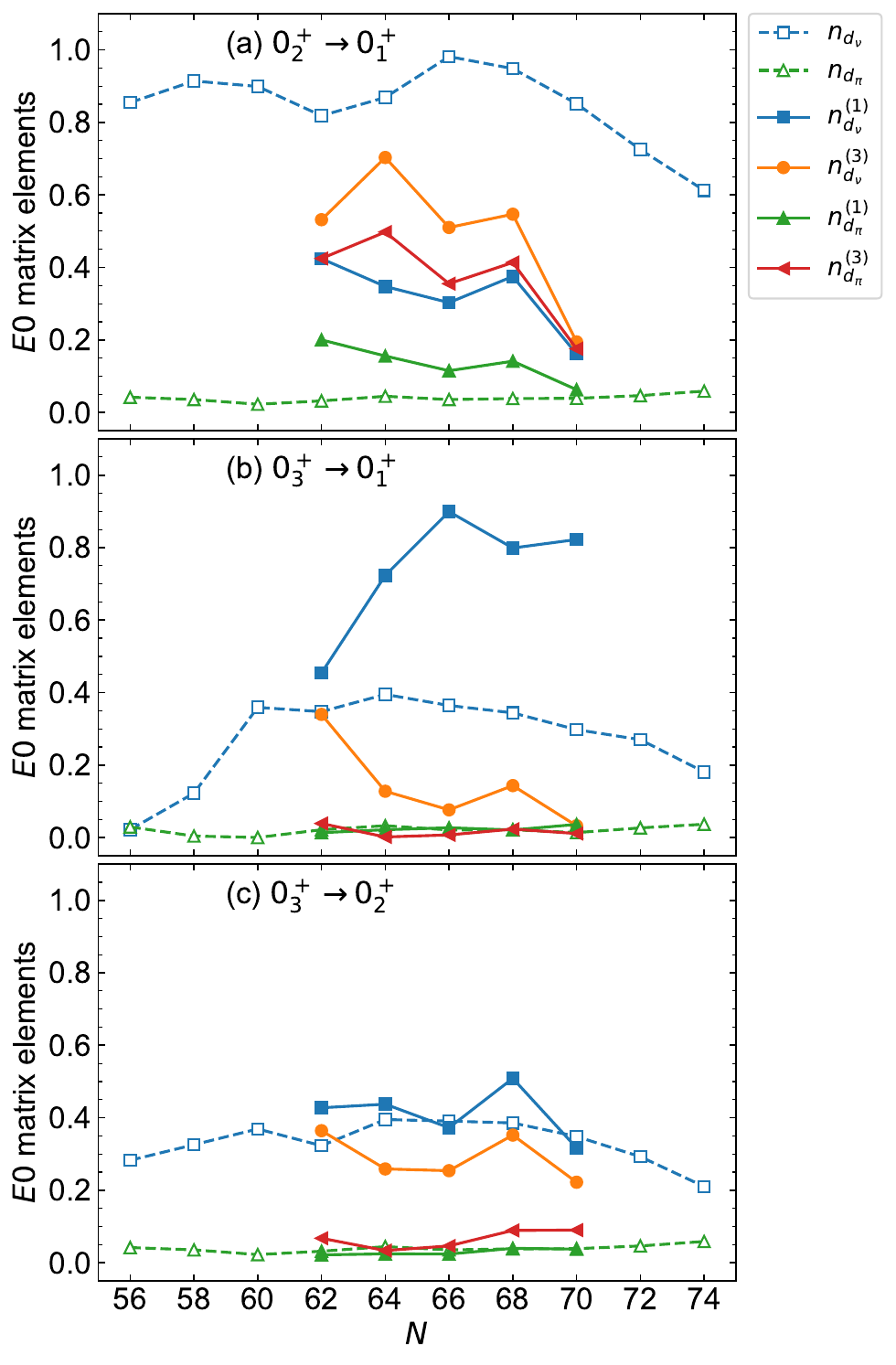}
\caption{
Matrix elements of terms in the
$E0$ operator for the IBM2-CM
and those for the IBM-2 for
the $\bej 0 2 0 1$, $\bej 0 3 0 1$,
and $\bej 0 3 0 2$ transitions
in Te isotopes.
See the main text for details.
}
\label{fig:e0-decom}
\end{center}
\end{figure}

Figure~\ref{fig:e0-decom} depicts
absolute values of
matrix elements of four terms
in the $E0$ operator
\eqref{eq:e0} denoted
$n^{(i)}_{d_{\rho}}(\bej 0 k 0 l)
\equiv
|\braket{
0^+_l|\hat P_i
\hat n_{d_{\nu}}
\hat P_i|0^+_k}|$,
which are
calculated for the
$\bej 0 2 0 1$, $\bej 0 3 0 1$,
and $\bej 0 3 0 2$ transitions
in the IBM2-CM.
These values are also
compared with the equivalent
matrix elements in the IBM-2,
defined as
$n_{d_{\rho}}(\bej 0 k 0 l)
\equiv
|\braket{
0^+_l|
\hat n_{d_{\rho}}
|0^+_k}|$.
It is seen from Fig.~\ref{fig:e0-decom}
that in the IBM-2
there is essentially no proton
contribution to all three
$E0$ transitions,
and that the neutron component
$n_{d_{\nu}}(\bej 0 2 0 1)$
is larger than
$n_{d_{\nu}}(\bej 0 3 0 1)$ and
$n_{d_{\nu}}(\bej 0 3 0 2)$ values
by a factor of 2-3.
In the IBM2-CM,
contributions from protons
for the normal and
intruder configurations,
$n_{d_{\pi}}^{(1,3)}(\bej 0 3 0 1)$
and
$n_{d_{\pi}}^{(1,3)}(\bej 0 3 0 2)$,
are basically negligible
for the $^{114-122}$Te nuclei.
The four components
$n_{d_{\rho}}^{(1,3)}$
appear to make sizable amounts of
contributions to the
$\bej 0 2 0 1$ transition
in the IBM2-CM.
It is of interest that
the neutron and proton
terms for the intruder
configurations,
$n_{d_{\rho}}^{(3)}(\bej 0 2 0 1)$,
make larger
contributions to the
$\bej 0 2 0 1$ transition
than those for the normal configuration,
$n_{d_{\rho}}^{(1)}(\bej 0 2 0 1)$,
and this finding implies that
configuration mixing
plays an important role
in the $\bej 0 2 0 1$ transition.
The $\bej 0 3 0 1$ transition is,
however, shown to be dominated
by the contribution from neutrons
in the normal configuration,
except for $^{114}$Te.
For the $^{114}$Te nucleus,
the neutron $d$-boson terms in the
intruder configuration also
makes a larger contribution
to the $\bej 0 3 0 1$ transition,
illustrating the
strong configuration mixing
in this nucleus in the present
calculation (see also Table~\ref{tab:wf}).
From Fig.~\ref{fig:e0-decom}(b)
it is seen that
the $\bej 0 3 0 2$ decay is
accounted for mainly by the
matrix elements
$n_{d_{\nu}}^{(1)}(\bej 0 3 0 2)$
and $n_{d_{\nu}}^{(3)}(\bej 0 3 0 2)$.
The fact that the latter
is of the same order of magnitude
as the former again indicates
importance of configuration
mixing in the $\bej 0 3 0 2$
transition.

To compare with
observable $E0$ properties,
e.g., $\rho^2(E0)$ values,
four parameters
$a_{\nu,1}$, $a_{\nu,3}$,
$a_{\pi,1}$, and $a_{\pi,3}$
in the $E0$ parameters need to be
determined.
It is, however, not obvious
to find optimal and
fixed values of
these parameters that can be used
for all the considered Te isotopes.
This is
partly because the experimental
information of the $E0$ transition
properties,
which can provide constraints
for the $E0$ parameters,
is not as abundant as in
the case of the neighboring
Sn and Cd nuclei.
As compared with the $E2$
and $M1$ transition properties,
the $E0$ transitions have not been
extensively studied in the IBM,
and the parameters for the IBM $E0$
operator appropriate
to reproduce the observed $E0$
properties have not been
very well known.
In addition, as shown
in Fig.~\ref{fig:e0-decom}
the calculated matrix elements
of the terms in the $E0$ operator
are rather sensitive
to the structure of the wave
functions of the initial and
final $0^+$ states.
The ambiguity of the $E0$
parameters was indeed
pointed out in the empirical IBM2-CM
calculation in Ref.~\cite{rikovska1989},
in which the $E0$ parameters
fitted to reproduce
the measured $\rho^2(E0)$
values for $^{122}$Te and $^{124}$Te
differ significantly.

%
%
\begin{figure}[ht]
\begin{center}
\includegraphics[width=\linewidth]{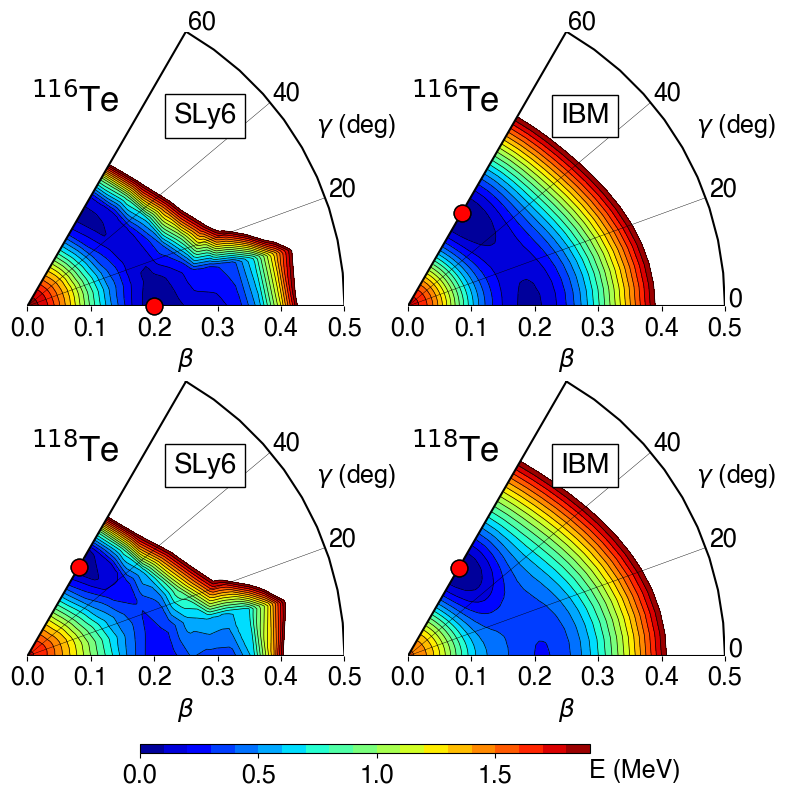}
\caption{
SCMF PESs for $^{116}$Te and $^{118}$Te
computed by the constrained HF+BCS method
using the Skyrme SLy6 EDF,
and the corresponding IBM2-CM PESs.
}
\label{fig:pes-sly6}
\end{center}
\end{figure}

%
%
\begin{figure*}[ht]
\begin{center}
\includegraphics[width=.35\linewidth]{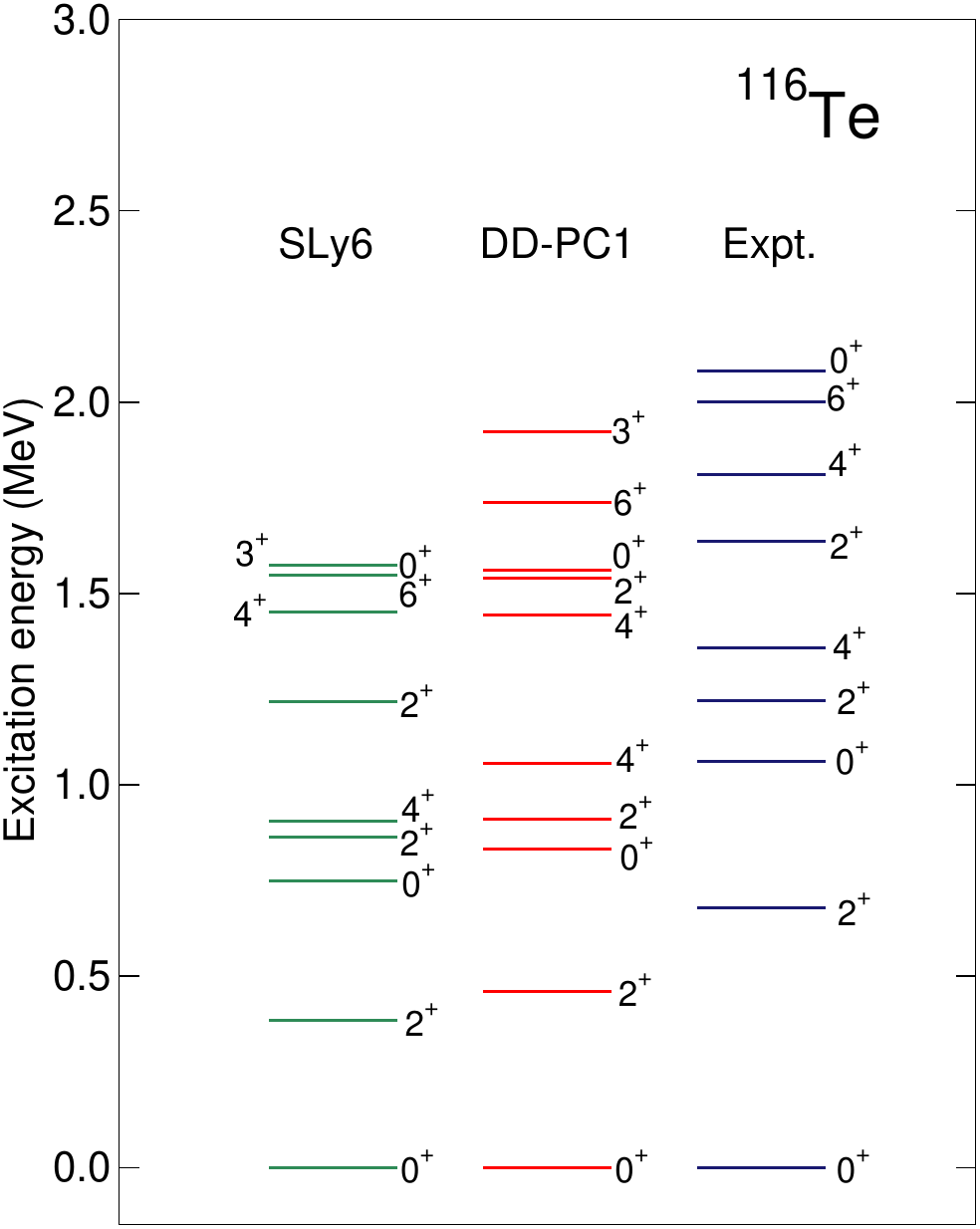}
\includegraphics[width=.35\linewidth]{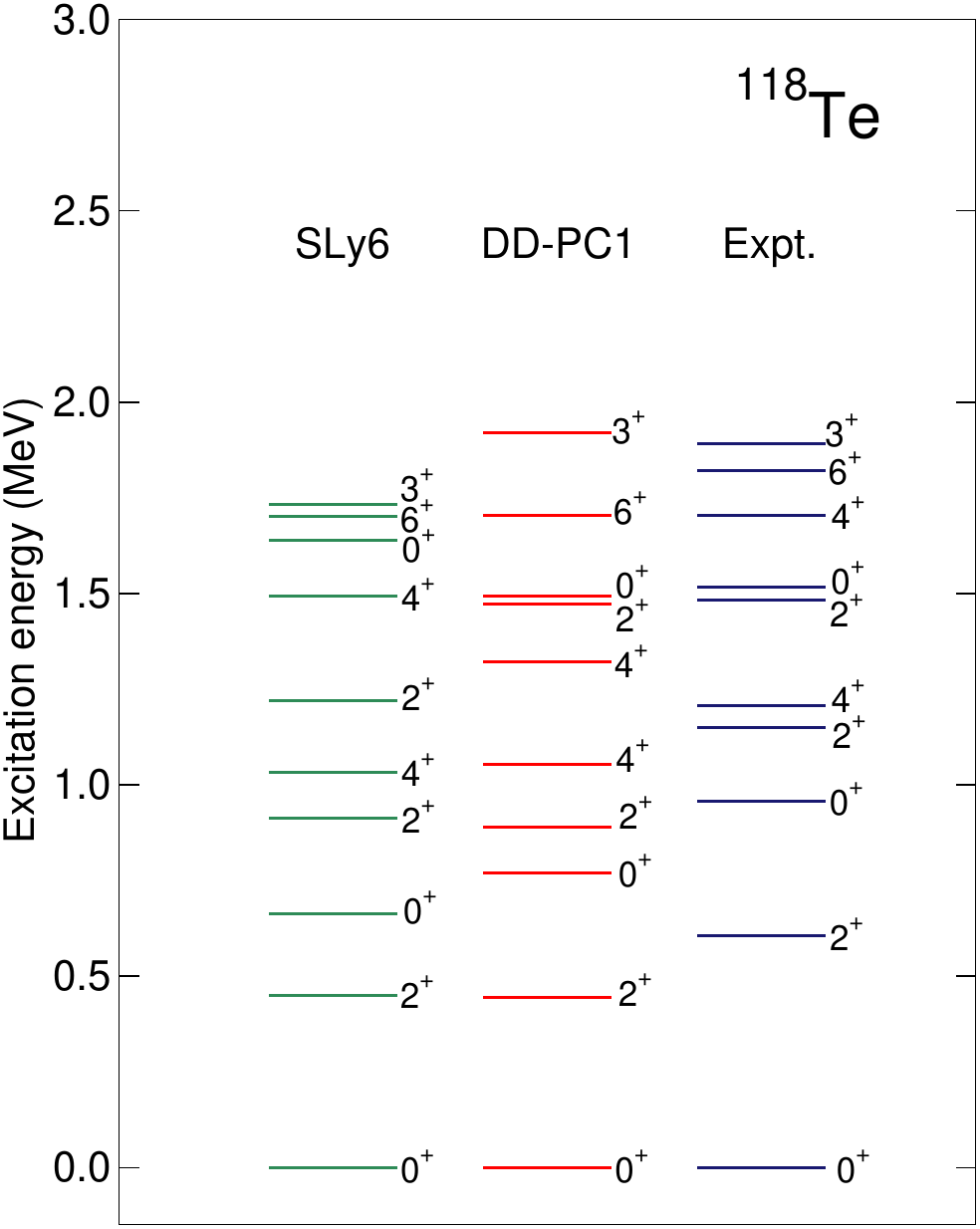}
\caption{
Comparisons of low-energy spectra
for $^{116}$Te and $^{118}$Te resulting
from the IBM2-CM mapping with
the Skyrme SLy6 force to those from
the DD-PC1 force in the
relativistic framework.
Experimental data are adopted from
Refs.~\cite{data,vonspee2024}.
}
\label{fig:level-sly6}
\end{center}
\end{figure*}

\subsection{Dependence on the EDF\label{sec:edf}}

As has been pointed out above,
even though many of the SCMF calculations
for the near midshell Te isotopes
have indicated the appearance of
prolate and oblate minima in
the energy surface,
the prolate-oblate energy balance
and $\gamma$ softness
seem to be at variance with
mean-field calculations
using different functionals.
Results of the present IBM2-CM calculations
for low-energy properties should,
therefore, depend on the choice of the EDF
that underlies the mapping procedure.

As an illustrative example,
Fig.~\ref{fig:pes-sly6} shows the
energy surfaces for $^{116}$Te
and $^{118}$Te obtained from the
constrained Hartree-Fock plus BCS (HF+BCS)
method \cite{ev8,ev8r}
using the Skyrme SLy6 EDF \cite{SLy}.
For the pairing channel,
the density-dependent zero-range
pairing force with the
strength of 1000 MeV fm$^3$
truncated below and above the
Fermi surface by 5 MeV for
both protons and neutrons is adopted.
For $^{116}$Te, the global
minimum is obtained on the prolate
side, which, however,
is only 14 keV in energy below the
minimum on the oblate side.
For $^{118}$Te, the prolate minimum
corresponds to the global minimum,
while there are two prolate local
minima.
Similarly to the RHB results
shown in Fig.~\ref{fig:pes1},
the HF+BCS calculations
predict for both nuclei
the coexistence
of prolate and oblate minima
in the PESs, but the Skyrme HF+BCS
energy surfaces are much
softer in the $\gamma$ deformation
than the RHB ones.
The mapped IBM2-CM PES for $^{116}$Te
reproduces the original HF+BCS PES,
but produces the global minimum
on the oblate side.
This is a consequence of the
configuration mixing
of the PESs for the two
unperturbed spaces \eqref{eq:pes},
which could
alter the energy of
the prolate minimum
to be slightly lower than
that of the prolate global minimum,
especially when the two
minima are quite close in energy.
As for $^{118}$Te, the normal
configuration is associated with
the oblate global minimum,
and the intruder configuration
with the prolate minimum
corresponding to the smaller
$\beta\approx0.2$ deformation.
The third minimum at $\beta\approx0.3$
could also have impacts on
the final results,
but is neglected here,
since the present calculation
is restricted to the mixing
of two configurations
and also for the sake of
simplicity.

In Fig.~\ref{fig:level-sly6},
the low-energy spectra
obtained from the IBM2-CM
using the Skyrme-SLy6
EDF are compared with those employing
the DD-PC1 EDF in the RHB calculations.
With the SLy6 EDF as a microscopic
input for the IBM2-CM,
the resultant energy spectrum
is rather compressed as
compared with the one
using the DD-PC1.
The reason for this
to occur is that the
present HF+BCS calculation
with the SLy6 EDF
suggests a steeper potential valley
along the $\beta$ deformation
in the PES, requiring
a larger quadrupole-quadrupole
strength, than the RHB with
the DD-PC1 force.
In addition, the $2^+_3$ and $3^+_1$
energy-levels in the Skyrme-SLy6
mapped IBM2-CM are particularly lower
than the DD-PC1 counterparts.
These states are supposed to
be members of $\gamma$-vibrational
bands in the present IBM2-CM
calculations, and are rather low in
energy when the Skyrme-SLy6 PESs
are considered as inputs,
which are $\gamma$ softer
than the DD-PC1 ones.

The degree of configuration mixing
is also more significant in the
Skyrme-SLy6 IBM2-CM than in the
case of the DD-PC1 EDF.
For instance, the intruder
prolate configuration accounts
for 48\% (30\%), 54\% (71\%), 50\% (71\%),
59\% (42\%), 61\% (62\%), 71\% (63\%)
of the wave functions for the
$0^+_1$, $0^+_2$, $0^+_3$,
$2^+_1$, $2^+_2$, and $4^+_1$
states for $^{116}$Te ($^{118}$Te),
respectively; these fractions
are closer to 50\% than for
the DD-PC1 results
(see Table~\ref{tab:wf}).

Tables~\ref{tab:te116-em} and
\ref{tab:te118-em} compare
the $B(E2)$ and $B(M1)$ values,
electric quadrupole and magnetic
dipole moments
resulting from the Skyrme-SLy6 mapped
IBM2-CM with those from the
DD-PC1 mapped IBM2-CM calculations.
The boson effective charges for
the $E2$ operator and effective
$g$ factors for the $M1$ operator
in the case of the SLy6 input
are chosen to be
the same as those for
the DD-PC1 case, but
an overall factor is multiplied
for the $E2$ operator
so that the
$\be 2 1 0 1$ values should
be equal to those from
the DD-PC1 RHB calculations.
The IBM2-CM with two different
microscopic inputs overall give
similar results.
There are also several
notable differences.
For instance,
the $\be 0 3 2 2$ values relative
to $\be 0 3 2 1$
are larger with the SLy6 input
than with the DD-PC1 input for
both nuclei.
Also, the sign $Q(2^+_1)<0$
in the SLy6-mapped IBM2-CM
for $^{116}$Te,
while it is positive
in the DD-PC1 case.
These differences between the
IBM2-CM calculations with the
different EDF inputs are attributed
to the softness in the
triaxial $\gamma$ deformation
of the SCMF PESs, since
the $\gamma$ softness
influences the degree of
the shape configuration
mixing.

\begin{table}
\caption{\label{tab:te116-em}
Calculated $B(E2)$ and $B(M1)$ values
(in W.u.), $Q(2^+_{1,2})$ (in $e$b),
and $\mu(2^+_1)$ (in $\mu_{N}$)
moments for $^{116}$Te obtained from
the IBM2-CM using the DD-PC1
and SLy6 EDFs as inputs.
Experimental data are taken
from Refs.~\cite{vonspee2024,pascu2025,data}.
}
 \begin{center}
 \begin{ruledtabular}
  \begin{tabular}{lcccc}
 & \multicolumn{2}{c}{IBM2-CM} & \multirow{2}{*}{Experiment}\\
\cline{2-3}
 & DD-PC1 & SLy6 & \\
\hline
$ B(E2; {2}^{+}_{1} \to {0}^{+}_{1})$ & $46$ & $46$ & $39^{+4}_{-3}$ \\ 
$ B(E2; {4}^{+}_{1} \to {2}^{+}_{1})$ & $76$ & $71$ & $52^{+6}_{-5}$ \\ 
$ B(E2; {2}^{+}_{2} \to {0}^{+}_{1})$ & $0.6$ & $0.09$ & ${}$ \\ 
$ B(E2; {2}^{+}_{2} \to {0}^{+}_{2})$ & $57$ & $14$ & ${}$ \\ 
$ B(E2; {2}^{+}_{2} \to {2}^{+}_{1})$ & $69$ & $65$ & $41^{+6}_{-5}$ \\ 
$ B(E2; {2}^{+}_{3} \to {0}^{+}_{1})$ & $0.09$ & $0.7$ & ${}$ \\ 
$ B(E2; {2}^{+}_{3} \to {0}^{+}_{2})$ & $23$ & $30$ & $>20$ \\ 
$ B(E2; {2}^{+}_{3} \to {2}^{+}_{1})$ & $7$ & $6$ & $>3$ \\ 
$ B(E2; {2}^{+}_{3} \to {2}^{+}_{2})$ & $7$ & $2$ & ${}$ \\ 
$ B(E2; {3}^{+}_{1} \to {2}^{+}_{2})$ & $28$ & $45$ & ${}$ \\ 
$ B(E2; {4}^{+}_{2} \to {2}^{+}_{1})$ & $0.1$ & $0.1$ & $>0.7$ \\ 
$ B(E2; {4}^{+}_{2} \to {2}^{+}_{2})$ & $80$ & $48$ & $>5$ \\ 
$ B(E2; {4}^{+}_{2} \to {4}^{+}_{1})$ & $59$ & $36$ & ${}$ \\ 
$ B(E2; {6}^{+}_{1} \to {4}^{+}_{1})$ & $108$ & $88$ & $67^{+9}_{-7}$ \\ 
$ B(E2; {0}^{+}_{2} \to {2}^{+}_{1})$ & $34$ & $27$ & $43^{+6}_{-5}$ \\ 
$ B(E2; {0}^{+}_{3} \to {2}^{+}_{1})$ & $14$ & $3$ & ${}$ \\ 
$ B(E2; {0}^{+}_{3} \to {2}^{+}_{2})$ & $14$ & $45$ & ${}$ \\ 
$ B(M1; {2}^{+}_{2} \to {2}^{+}_{1})$ & $3.60\times10^{-4}$ & $2.64\times10^{-4}$ & $6^{+6}_{-5}\times10^{-4}$ \\ 
$ B(M1; {2}^{+}_{3} \to {2}^{+}_{1})$ & $4.20\times10^{-5}$ & $3.88\times10^{-4}$ & $>0.01$ \\ 
$ B(M1; {2}^{+}_{3} \to {2}^{+}_{2})$ & $2.82\times10^{-4}$ & $2.00\times10^{-6}$ & ${}$ \\
$ B(M1; {3}^{+}_{1} \to {2}^{+}_{1})$ & $1.20\times10^{-5}$ & $1.00\times10^{-6}$ & ${}$ \\ 
$ B(M1; {3}^{+}_{1} \to {2}^{+}_{2})$ & $1.53\times10^{-4}$ & $3.60\times10^{-5}$ & ${}$ \\ 
$ B(M1; {3}^{+}_{1} \to {4}^{+}_{1})$ & $1.27\times10^{-4}$ & $3.20\times10^{-5}$ & ${}$ \\ 
$ B(M1; {4}^{+}_{2} \to {4}^{+}_{1})$ & $1.77\times10^{-3}$ & $1.03\times10^{-3}$ & ${}$ \\ 
$ Q({2}^{+}_{1})$ & $0.14$ & $-0.10$ & ${}$ \\ 
$ Q({2}^{+}_{2})$ & $-0.67$ & $-0.14$ & ${}$ \\
$\mu({2}^{+}_{1})$ & $0.54$ & $0.56$ & ${}$ \\
 \end{tabular}
 \end{ruledtabular}
 \end{center}
\end{table}

\begin{table}
\caption{\label{tab:te118-em}
Same as the caption to Table~\ref{tab:te116-em},
but for $^{118}$Te.
Experimental data are taken
from Refs.~\cite{cederlof2023-118Te,mihai2011,data}.
}
 \begin{center}
 \begin{ruledtabular}
  \begin{tabular}{lcccc}
 & \multicolumn{2}{c}{IBM2-CM} & \multirow{2}{*}{Experiment}\\
\cline{2-3}
 & DD-PC1 & SLy6 & \\
\hline
$ B(E2; {2}^{+}_{1} \to {0}^{+}_{1})$ & $45$ & $45$ & $38.93\pm0.98$ \\ 
$ B(E2; {4}^{+}_{1} \to {2}^{+}_{1})$ & $71$ & $74$ & $71.3\pm3.8$ \\ 
$ B(E2; {2}^{+}_{2} \to {0}^{+}_{1})$ & $1.0$ & $0.4$ & $1.8^{+0.7}_{-0.4}$ \\ 
$ B(E2; {2}^{+}_{2} \to {0}^{+}_{2})$ & $87$ & $38$ & ${}$ \\ 
$ B(E2; {2}^{+}_{2} \to {2}^{+}_{1})$ & $39$ & $51$ & $28^{+13}_{-7}$\footnotemark[1] \\ 
& & & $264^{+100}_{-57}$\footnotemark[2] \\ 
$ B(E2; {2}^{+}_{3} \to {0}^{+}_{1})$ & $0.1$ & $0.07$ & $0.20^{+0.11}_{-0.05}$ \\ 
$ B(E2; {2}^{+}_{3} \to {0}^{+}_{2})$ & $16$ & $21$ & $60^{+35}_{-17}$ \\ 
$ B(E2; {2}^{+}_{3} \to {2}^{+}_{1})$ & $15$ & $23$ & $3.4^{+1.9}_{-0.9}$\footnotemark[3] \\ 
& & & $30^{+16}_{-7}$\footnotemark[4] \\ 
$ B(E2; {2}^{+}_{3} \to {2}^{+}_{2})$ & $3$ & $0.9$ & $9.5^{+8.5}_{-3.5}$\footnotemark[5] \\ 
& & & $166^{+119}_{-61}$\footnotemark[6] \\
$ B(E2; {3}^{+}_{1} \to {2}^{+}_{1})$ & $0.7$ & $0.03$ & $<0.036$ \\ 
$ B(E2; {3}^{+}_{1} \to {2}^{+}_{2})$ & $20$ & $29$ & $<0.7$ \\ 
$ B(E2; {3}^{+}_{1} \to {4}^{+}_{1})$ & $17$ & $26$ & $<4.5$ \\ 
$ B(E2; {4}^{+}_{2} \to {2}^{+}_{1})$ & $0.7$ & $0.01$ & $<1.1$ \\ 
$ B(E2; {4}^{+}_{2} \to {2}^{+}_{2})$ & $116$ & $65$ & $<210$ \\ 
$ B(E2; {4}^{+}_{2} \to {4}^{+}_{1})$ & $59$ & $32$ & $<160$ \\ 
$ B(E2; {6}^{+}_{1} \to {4}^{+}_{1})$ & $96$ & $101$ & $80^{+14}_{-10}$ \\ 
$ B(E2; {0}^{+}_{2} \to {2}^{+}_{1})$ & $14$ & $33$ & $54\pm45$ \\ 
$ B(E2; {0}^{+}_{3} \to {2}^{+}_{1})$ & $22$ & $10$ & $1.3\pm0.2$ \\ 
$ B(E2; {0}^{+}_{3} \to {2}^{+}_{2})$ & $5$ & $15$ & $100$ \\
$ B(M1; {2}^{+}_{2} \to {2}^{+}_{1})$ & $1.40\times10^{-4}$ & $3.68\times10^{-4}$ & $9.5^{+3.5}_{-2.1}\times10^{-2}$\footnotemark[1] \\ 
 & & & $8.5^{+3.5}_{-2.8}\times10^{-4}$\footnotemark[2] \\ 
$ B(M1; {2}^{+}_{3} \to {2}^{+}_{1})$ & $2.00\times10^{-6}$ & $1.43\times10^{-4}$ & $2.8^{+1.4}_{-0.6}\times10^{-2}$\footnotemark[3] \\ 
& & & $1.3^{+0.6}_{-0.5}\times10^{-4}$\footnotemark[4] \\ 
$ B(M1; {2}^{+}_{3} \to {2}^{+}_{2})$ & $1.69\times10^{-4}$ & $2.10\times10^{-5}$ & $2.4^{+1.8}_{-0.9}\times10^{-2}$\footnotemark[5] \\ 
& & & $1.2^{+0.9}_{-0.5}\times10^{-3}$\footnotemark[6] \\
$ B(M1; {3}^{+}_{1} \to {2}^{+}_{1})$ & $3.00\times10^{-6}$ & $1.84\times10^{-7}$ & $<1.6\times10^{-3}$ \\ 
$ B(M1; {3}^{+}_{1} \to {2}^{+}_{2})$ & $1.26\times10^{-4}$ & $1.80\times10^{-5}$ & $<1.8\times10^{-2}$ \\ 
$ B(M1; {3}^{+}_{1} \to {4}^{+}_{1})$ & $6.50\times10^{-5}$ & $2.50\times10^{-5}$ & $<1\times10^{-2}$ \\ 
$ B(M1; {4}^{+}_{2} \to {4}^{+}_{1})$ & $1.20\times10^{-3}$ & $1.40\times10^{-3}$ & $<0.018$ \\ 
$ Q({2}^{+}_{1})$ & $0.59$ & $0.12$ & ${}$ \\ 
$ Q({2}^{+}_{2})$ & $-1.22$ & $-0.46$ & ${}$ \\
$\mu({2}^{+}_{1})$ & $0.52$ & $0.55$ & ${}$ \\
 \end{tabular}
 \end{ruledtabular}
\footnotetext[1]
{Based on the mixing ratio $\delta=-0.35\pm0.02$.}
\footnotetext[2]
{Based on the mixing ratio $\delta=11.0^{+0.7}_{-0.5}$.}
\footnotetext[3]
{Based on the mixing ratio $\delta=-0.36\pm0.02$.}
\footnotetext[4]
{Based on the mixing ratio $\delta=16.0\pm2$.}
\footnotetext[5]
{Based on the mixing ratio $\delta=-0.25\pm0.5$.}
\footnotetext[6]
{Based on the mixing ratio $\delta=4.4\pm0.3$.}
 \end{center}
\end{table}

\section{Summary and conclusions\label{sec:summary}}

Evolution and coexistence of the
intrinsic shapes and related low-energy
spectroscopic properties in the
even-even Te isotopes have been investigated.
The IBM2-CM has been employed to
distinguish possible intruder states
from the low-lying states in midshell nuclei
$^{114-122}$Te.
The interaction strengths in the
Hamiltonian of the IBM2-CM have been
determined for each nucleus
by mapping the RHB-SCMF PES
with a given EDF and a pairing
interaction onto the
equivalent energy surface of the
IBM2-CM.
From the systematic of the
calculated PESs the transition
from the prolate to oblate shapes
was suggested to occur at
$^{114}$Te, and a pronounced
oblate-prolate shape coexistence
was found in $^{116,118,120}$Te.

The calculated low-energy levels
for the $0^+_2$, $2^+_2$, $0^+_3$,
and $2^+_3$ states
exhibit parabolic behaviors with the
lowest energies at the midneutron-shell
nucleus $^{118}$Te,
which behaviors are characteristics
of the emergent shape-existence structure
observed in the neighboring Sn and
Cd isotopic chains.
To reproduce the observed
systematic of the
nonyrast energy levels,
it appears to be reasonable to
take into account the configuration
mixing in the IBM framework.
In many of the low-energy and low-spin
states particularly in the $^{114-120}$Te
nuclei, the intruder configurations
that are associated with the prolate shape
having a large quadrupole deformation
were shown to play an important role.
Large predicted $B(E2)$ values
such as $\be 0 2 2 1$ and
$\be 2 2 2 1$, particularly at
$^{114}$Te and $^{116}$Te,
are also signatures
of configuration mixing.
The predicted low-energy level
structure and $E2$ transitions
in the midshell region
suggest a band built on the $0^+_2$
state with a possible member being
the $2^+_2$ or $2^+_3$ state.
The $0^+_2$ states in the present
calculations were shown to be
characterized by large amounts of
the intruder configurations.
Concerning the $E2$ transitions
of the yrast states, however,
in some cases
the standard IBM-2 description
with only a single
configuration was shown to
be adequate to reasonably
reproduce the experimental data
in comparison with the IBM2-CM.
The inclusion of the configuration
mixing, therefore, appears not to
lead to complete descriptions
of these transition properties,
since the intruder configuration
is to a great
extent included in the
corresponding wave functions
and the admixture of the
normal and intruder configurations
makes it less obvious
to properly account for
the expected transition patterns
in detail.

In particular, difficulties arose
in reproducing
the anomalous
reduction of the
$\be 4 1 2 1$ and $\be 6 1 4 1$
values at $^{114}$Te,
as observed experimentally.
Neither of the present IBM2-CM
and standard IBM-2,
which are based on
the microscopic EDF,
and conventional IBM2-CM
with parameters fitted to data,
was not able to account for
that phenomenon.
The local systematic of
these $E2$ transitions thus
suggests that certain extensions
of the IBM framework
in general are in order
for realistic nuclear
structure studies.
This will be a challenging,
but interesting future study
for a further development
of the IBM2-CM that
is based on the EDF.
Furthermore, the calculated
spectroscopic quadrupole
moments suggested that the
$2^+_1$ states in heavier Te
were oblate in nature,
which is inconsistent with
experiment.
These results indicate
that the structure of the $2^+_1$
states may not be
correctly described.
The spectroscopic results
of the IBM2-CM are also influenced
by a particular choice of the
underlying EDF, since especially
the prolate-oblate energy balance
in the PES depends largely
on the properties of the EDF.
In some particular cases,
the IBM2-CM combined with
a given choice of the EDF
may not be adequate to
describe very accurately
the observed spectroscopic
properties.
Certain improvements of the
methodology are in order
at the SCMF and/or IBM levels.

The results obtained
from the present
IBM2-CM calculations
indicate the relevance of the intruder
configurations and their couplings to
the low-lying states in the mid-neutron-shell
and neutron-deficient nuclei
in the region with $Z>50$.
The present theoretical analysis
can be extended to provide
predictions on nuclear
spectroscopy related to
shape coexistence in other mass
regions, which have been
previously beyond reach of
microscopic nuclear structure models.
This will be a step toward
establishing a comprehensive
picture of the shape coexistence
in nuclei.

\acknowledgments
This work has been supported
by JSPS KAKENHI Grant No. JP25K07293.

\bibliography{refs}

\end{document}